\documentclass[10.5pt,groupedaddress, showpacs, superscriptaddress]{revtex4-2}
\usepackage{amssymb,amsmath,amsthm,color,graphicx,times,graphicx}
\graphicspath{{./figures/}}
\usepackage{hyperref}
\usepackage{subfigure}
\usepackage{times}
\usepackage{bbold}
\usepackage{color}
\usepackage{array}
\usepackage{sidecap}
\usepackage[french]{babel}
\usepackage{array,multirow,makecell}
\usepackage[table]{xcolor}

\usepackage{orcidlink}
\theoremstyle{plain}

\theoremstyle{definition}

\begin{document}
\title{Achieving quantum metrological performance and exact Heisenberg limit precision through superposition of $s$-spin coherent states}

\author{Hanan Saidi}\affiliation{LPHE-Modeling and Simulation, Faculty of Sciences, Mohammed V University in Rabat, Rabat, Morocco.}
\author{Hanane El Hadfi}\affiliation{LPHE-Modeling and Simulation, Faculty of Sciences, Mohammed V University in Rabat, Rabat, Morocco.}\affiliation{Centre of Physics and Mathematics, CPM, Faculty of Sciences, Mohammed V University in Rabat, Rabat, Morocco.}
\author{Abdallah Slaoui \orcidlink{0000-0002-5284-3240}}\email{Corresponding author: abdallah.slaoui@um5s.net.ma}\affiliation{LPHE-Modeling and Simulation, Faculty of Sciences, Mohammed V University in Rabat, Rabat, Morocco.}\affiliation{Centre of Physics and Mathematics, CPM, Faculty of Sciences, Mohammed V University in Rabat, Rabat, Morocco.}\affiliation{Center of Excellence in Quantum and Intelligent Computing, Prince Sultan University, Riyadh, Saudi Arabia.}
\author{Rachid Ahl Laamara}\affiliation{LPHE-Modeling and Simulation, Faculty of Sciences, Mohammed V University in Rabat, Rabat, Morocco.}\affiliation{Centre of Physics and Mathematics, CPM, Faculty of Sciences, Mohammed V University in Rabat, Rabat, Morocco.}

\begin{abstract}
In quantum phase estimation, the Heisenberg limit provides the ultimate accuracy over quasi-classical estimation procedures. However, realizing this limit hinges upon both the detection strategy employed for output measurements and the characteristics of the input states. This study delves into quantum phase estimation using $s$-spin coherent states superposition. Initially, we delve into the explicit formulation of spin coherent states for a spin $s=3/2$. Both the quantum Fisher information and the quantum Cramer–Rao bound are meticulously examined. We analytically show that the ultimate measurement precision of spin cat states approaches the Heisenberg limit, where uncertainty decreases inversely with the total particle number. Moreover, we investigate the phase sensitivity introduced through operators $e^{i\zeta{S}_{z}}$, $e^{i\zeta{S}_{x}}$ and $e^{i\zeta{S}_{y}}$, subsequently comparing the resultants findings. In closing, we provide a general analytical expression for the quantum Cramér–Rao bound applied to these three parameter-generating operators, utilizing general $s$-spin coherent states. We remarked that attaining Heisenberg-limit precision requires the careful adjustment of insightful information about the geometry of $s$-spin cat states on the Bloch sphere. Additionally, as the number of $s$-spin increases, the Heisenberg limit decreases, and this reduction is inversely proportional to the $s$-spin number.

\end{abstract}
\date{\today}

\maketitle
\section{ Introduction}
Quantum metrology holds a pivotal role across a spectrum of technological domains \cite{Nielsen2001,Bellac2006,Simon2017,Barbieri2022,Dakir2023,Polino2020,Friis2017,Taylor2016,Schnabel2010}. Within the realm of quantum information, metrology strives to achieve heightened precision measurements utilizing quantum resources in quantum states such as coherent states, entangled states, and squeezed states \cite{Przysiezna2015,Huang2014}. Its primary objective lies in refining the precision available for estimating physical parameters and transcending the confines of the standard quantum limit. This field has illuminated the potential of quantum metrology in devising measurement techniques that outperform classical counterparts in terms of precision \cite{Bakraoui2022,Nada2023,Mitchell2004}. As a consequence, leveraging quantum effects like quantum entanglement leads to a reduction in fluctuations by a factor proportional to $1/N$, referred to as the Heisenberg limit. In essence, the Heisenberg limit defines the ultimate precision threshold attainable through measurement \cite{Nagata2007,Abouelkhir2023A}. Consequently, advancements in precision can be realized through the utilization of exquisitely sensitive interferometric setups, such as the Mach-Zehnder interferometer involving $N$ entangled particles \cite{Slaoui2022,Muller2022}. Furthermore, non-classical states like entangled coherent states (ECS) and generalized NOON states exhibit analogous enhancements in accuracy \cite{Zwierz2010,Holland1993,Pezze2002,Joo2011}. The Heisenberg limit can similarly be approached through the application of spin squeezed states and spin coherent states \cite{Hayashi2006,Maleki2021}. Spin coherent states emerge as crucial entities in quantum metrology, offering a means to measure physical quantities with superior precision compared to classical methodologies. These states confer numerous benefits to quantum metrology, including heightened sensitivity to external fields, diminished measurement noise, and the potential for meticulous preparation with high coherence \cite{Berrada2012,Zhang2014,Maleki2020}. These advantages collectively render spin coherent states a valuable asset for the precise measurement of diverse physical quantities across a wide array of applications.\par

Although the Heisenberg Limit defines the ultimate limit for measurement precision, the practical accuracy achieved in a specific experiment hinges on the properties of the chosen quantum probes. Among these probes, NOON states stand out for enabling HL-like precision \cite{Giovannetti2011,Dowling2008,Boto2000}. However, generating NOON states presents significant conceptual and technical challenges. While recent years have seen the development of efficient schemes using tailored quantum-state engineering to address these difficulties (see Refs.\cite{Riedel2010,Maleki22018,Su2017,Maleki22019}), the quest continues for practical parameter-estimation solutions that utilize a broader class of superposition coherent states. This broader approach would not only facilitate experimental studies but also help tackle issues like decoherence, dissipation, and the high cost and inefficiency of current state-generation methods. A key challenge remains that of achieving the Heisenberg Limit in the presence of noise. Particle loss is a major culprit, as it rapidly degrades phase information in the resulting mixed state. Recent research has explored the potential benefits of nonlinearities and the critical role of query complexity in quantum metrology \cite{Rivas2010}. This concept focuses on the resource disparity between different states needed for the same phase estimation task \cite{Zwierz2010}. On the other hand, researchers have proposed various theoretical and experimental strategies to achieve high-resolution measurements \cite{Toth2014,Liu2022}. These strategies fall into two main categories: those using classical states with nonlinear Hamiltonians and those using nonclassical states with linear Hamiltonians. One approach involves reaching the measurement limit using separable states under a nonlinear Hamiltonian \cite{Boixo2008}. Alternatively, maximally entangled states constructed under a linear Hamiltonian can also achieve this limit \cite{Roy2008,Boixo2007}.\par

Quantum metrology with coherent state superposition represents an advanced technique that enables more precise measurements compared to those achieved using simple coherent states \cite{Birrittella2021,Huang2018}. This method leverages the quantum properties of particles to make exceedingly accurate measurements. By employing spin coherent state superposition, it becomes feasible to ascertain magnetic fields with heightened precision compared to classical techniques. Similarly, these states find utility in quantifying phase parameters, such as rotation angles or polarizations. Within the realm of estimation theory, resolving a parameter estimation challenge entails the discovery of an estimator, denoted as $\hat{\zeta}$. This estimator functions as an operation that furnishes a collection of measurement outcomes within the parameter space. Furthermore, the attainment of utmost precision hinges on the quantum Fisher information, a critical factor for establishing the quantum Cramér–Rao bound \cite{Helstrom1969,Paris2009}. The most effective estimators are those that reach the limit set by this Cramer-Rao inequality:
\begin{equation}
	{\rm Var}(\hat{\zeta})\geq\frac{1}{nF(\zeta)},
	\label{1}
\end{equation}
where ${\rm Var}(\cdot)$ represents the variance, $n$ denotes the number of repeated experiments, $\hat{\zeta}$ is the estimator for the parameter $\zeta$, and $F(\zeta)$ stands for the Quantum Fisher Information (QFI). The QFI of a state $\rho_{\zeta}$ is defined as \cite{Helstrom1969,Paris2009}
\begin{equation}
	F(\zeta)=Tr[\rho_{\zeta} L_{\zeta}^{2}].
	\label{2}
\end{equation}
Here, the operator denoted as $L_{\zeta}$ is commonly referred to as the symmetric logarithmic derivative operator, and its explicit determination is established through the following relation
\begin{equation}
	\frac{\partial{\rho}_{\zeta}}{\partial{\zeta}}=\frac{{L_{\zeta}}{\rho_{\zeta}}+{\rho_{\zeta}}{L_{\zeta}}}{2}.
	\label{3}
\end{equation}
Certainly, the QFI plays a pivotal role within the realm of quantum metrology. Its utility extends to extracting insights about parameters that are not directly measurable, as indicated in \cite{Paris2009}. Notably, this metric serves as a gauge for quantifying the phase sensitivity of quantum systems, as demonstrated in \cite{Yu2018,Abouelkhir2023,Yi2012}. Importantly, it's worth noting that this metric can be directly derived when the explicit calculation of the symmetric logarithmic derivative operator is performed. For mixed states \cite{Jing2014,SlaouiB2019}, the QFI can be expressed by utilizing the spectral decomposition of the output state \(\rho_{\zeta}=\sum_{i=1}^{s}p_{i}{{|\Psi_{i}\rangle}{\langle{\Psi_{i}|}}}\), where $s$, $p_{i}$, and $|{\Psi_{i}}\rangle$ represent the dimension of the support for the matrix, the eigenvalues, and the eigenstates of $\rho_{\zeta}$, respectively. Then the QFI can be expressed by 
\begin{equation}
	F(\zeta)=\sum_{i=1}^{s}{\frac{1}{p_{i}}}(\partial_{\zeta}p_{i})^{2}+{\sum_{i=1}^{s}{4p_{i}\langle{\partial_{\zeta}\Psi_{i}|\partial_{\zeta}\Psi_{i}}\rangle}}-{\sum_{i,j=1}^{s}{\frac{8p_{i}p_{j}}{p_{i}+p_{j}}}{|\langle{\Psi_{i}|\partial_{\zeta}\Psi_{j}}\rangle}|^{2}}.
	\label{4}
\end{equation}
For the unitary parameterization process $\rho_{\zeta}=e^{-i{\zeta}H}{\rho}e^{+i{\zeta}H}$, with $H$ a Hermitian operator, the expression of the QFI reduces to
\begin{equation}
	F(\zeta)={\sum_{i=1}^{s}{4p_{i}\langle{\Psi_{i}|H^{2}|\Psi_{i}}\rangle}}-{\sum_{i,j=1}^{s}{\frac{8p_{i}p_{j}}{p_{i}+p_{j}}}{|\langle{\Psi_{i}|H|\Psi_{j}}\rangle}|^{2}}.
	\label{5}
\end{equation}
Furthermore, for pure quantum states $\rho_{\zeta}= |{\Psi_{\zeta}}\rangle{\langle}\Psi_{\zeta}|$, the QFI is given by
\begin{equation}
	F(\rho_{\zeta})=4[\langle{\Psi}|H^{2}|{\Psi}\rangle-|\langle{\Psi}|H|{\Psi\rangle}|^{2}]=4\Delta^{2}H,
	\label{6}
\end{equation}
where $H$ is the Hamiltonian that generates the interaction between the probe and the object being measured.\par 
In this paper, we focus on the phase estimation protocol in quantum metrology, specifically centered around the task of estimating an unknown parameter $\zeta$ in the context of a $3/2$-spin system that involves the superposition of spin coherent states. We commence our exploration in Section (\ref{a}) by establishing the groundwork through the introduction of spin coherent states. Subsequently, in Section (\ref{b}), we delve into the calculation of both the quantum Fisher information and the quantum Cramer–Rao bound. We initiate our calculations by determining the quantum Fisher information and the quantum Cramer–Rao bound for scenarios wherein the system accumulates a phase via operations $e^{i\zeta{S}_{z}}$, $e^{i\zeta{S}_{x}}$, and $e^{i\zeta{S}_{y}}$. Specifically, we explore circumstances where optimal estimation of the phase shift parameter can be attained. Additionally, we showcase the substantial enhancement in measurement precision achievable through the utilization of a superposition of spin coherent states. Moving forward to Section (\ref{x}), we offer a comprehensive analytical expression for quantum Fisher information linked to the Cramér–Rao bound of an unknown parameter, applicable to coherent states of any $s$-spin. Finally, our insights culminate in Section (\ref{y}) where we provide our concluding remarks.
\section{A brief review on spin coherent states}\label{a}
Spin coherent states play a pivotal role in the characterization and utilization of the quantum properties exhibited by spin systems. These distinctive states, often labeled as Bloch states, embody a coherent superposition of diverse spin states within Hilbert space \cite{Monroe1996,Klauder1985,Zhang1990}. They furnish an elegant and potent mathematical portrayal of a particle's spin orientation, thereby facilitating a deeper comprehension of the quantum intricacies underlying spin precession and evolution. Notably, the hallmark of spin-coherent states lies in their capacity to minimize the uncertainty inherent in measuring spin orientation. Their coherently structured essence engenders distributional attributes that evenly encompass the entirety of the Bloch sphere, bestowing heightened sensitivity to even the slightest deviations in spin angles. This feature makes them exceptionally suitable for applications in the realm of quantum metrology, where high-precision measurements are needed to accurately estimate unknown parameters. Broadly, coherent states exhibit numerous intriguing properties that have positioned them as the "classical states" of the harmonic oscillator \cite{Glauber1963,Perelomov1972,Gazeau2009}. The fundamental goals of this study encompass the exploration of spin coherent states (SU(2) coherent states) and their associated quantum metrological capabilities. Here, we elucidate the key characteristics inherent to spin coherent states. These states \cite{Agarwal2013}, alternatively referred to as Bloch coherent states or atomic states, are defined by
\begin{equation}
	\begin{split}
		|\theta,\varphi,s\rangle&=R(\theta,\varphi)|s,-s\rangle\\&=\exp{\lbrack}\frac{-\theta}{2}(S_{+}e^{-i{\varphi}}-S_{-} e^{i{\varphi}})\rbrack|s,-s\rangle\\&=\sum_{m=-s}^{s} C_{m}(\theta)e^{-{i(s+m)\varphi}}|s,m\rangle.
		\label{8}
	\end{split}
\end{equation}
with ${C_{m}(\theta)=\sqrt{\frac{2s!}{(s+m)! (s-m)!}}{\cos^{s-m}({\frac{\theta}{2}})}\sin^{s+m}({\frac{\theta}{2}})}$, ${|s,m\rangle}$ denotes the Dicke state; symmetric superpositions of states involving $N$ spins, with $s+m$ particles in one state and $s-m$ particles in the other, where $s = N/2$, $m = -s,-s+ 1,...,s-1,s$, and $R(\theta,\varphi)$ represents the rotation operator. The overlap between two spin coherent states is given by
\begin{equation}
\langle{\theta_{1},\varphi_{1},s}|{\theta_{2},\varphi_{2},s}\rangle=\Bigl(\cos{(\frac{\theta_{1}}{2})}\cos{(\frac{\theta_{2}}{2})}+e^{i({\varphi_{1}}-{{\varphi_{2}}})}\sin{(\frac{\theta_{1}}{2})}\sin{(\frac{\theta_{2}}{2})}\Bigl)^{2s}.
\label{12}
\end{equation}
The parameters $\varphi$ and $\theta$ are, respectively, the azimuthal and polar coordinates on the Bloch sphere, $S_{-}$ and $S_{+}$ are  often called lowering and raising operators. A collection of $N$ two-state bosons, symmetrical by particle exchange, is analogous to a system with $N/2$ angular momentum. Within this framework, the link between the basis of two-mode bosonic harmonic oscillators and the Dicke basis becomes apparent through an examination of the Schwinger representation of the $SU(2)$ algebra within the context of the two-mode Hilbert space, given by
\begin{equation}
S_{+}=\hat{a}^{\dagger}\hat{b}, \hspace{1cm}S_{-}=\hat{b}^{\dagger}\hat{a},\hspace{1cm}S_{z}=\frac{1}{2}\left(\hat{a}^{\dagger}\hat{a}-\hat{b}^{\dagger}\hat{b}\right), 
\end{equation}
satisfying the commutation relations
\begin{equation}
	\lbrack{S_{+},S_{-}}\rbrack=2S_{z},\hspace{1cm} \lbrack{S_{z},S_{\pm}}\rbrack={\pm}S_{\pm},
	\label{9}
\end{equation}
where $\hat{a}$ and $\hat{b}$ denote the annihilation operators corresponding to the first and second modes of the bosonic harmonic oscillator. This realization, which finds applicability in diverse scenarios, leads us to refer to these operators as the photon annihilation operators. When the combined photon count in the two modes $\hat{a}$ and $\hat{b}$ totals $N$, with $s=N/2$ and $m=\left(N_{a}-N_{b}\right)/2$, the Dicke basis can be represented as $\left|N_{a}\right\rangle\left|N_{b}\right\rangle\equiv \left|s,m\right\rangle$ (with $N=N_{a}+N_{b}$). These generators act on an irreducible unitary representation as follows
\begin{equation}
	S_{\pm}|s,m\rangle=\sqrt{s(s+m)-m(m{\pm}1)}|s,m{\pm}1\rangle,\hspace{1cm} S_{z}|s,m\rangle=m|s,m\rangle,
	\label{10}
\end{equation}
and the Casimir operator is the quadratic operator $S^{2}$ which commutes with the operators $S_{\pm}$ and $S_{z}$. From $S_{\pm}$ ($S_{\pm}=S_{x}\pm iS_{y}$), the self-adjoint operator can be defined as
\begin{equation}
	S_{x}=\frac{S_{+}+S_{-}}{2},\hspace{1cm} S_{y}=\frac{S_{+}-S_{-}}{2i}.
	\label{11}
\end{equation}
Besides, in the case where all photons are confined to a single mode, we encounter two distinct scenarios: $|0\rangle\otimes|N\rangle$ (equivalent to $|s,-s\rangle$) and $|N\rangle\otimes|0\rangle$ (equivalent to $|s,s\rangle$). To emphasize the importance of spin-coherent states in quantum metrology, it's worth pointing out that the superposition of these two special cases yields the N00N state in the form of
$|NOON\rangle=\left(|N, 0\rangle + |0,N\rangle\right)/\sqrt{2}$. Hence, within the Dicke basis, the NOON state can be interpreted as a superposition encompassing the north and south poles on the Bloch sphere as $|NOON\rangle=\left(|s, s\rangle+|s, -s\rangle\right)/\sqrt{2}$. These states represent a distinct category of entangled quantum states that enhance precision in the estimation of physical parameters and enable measurements to be made beyond the standard quantum limits \cite{Huang2015,Sanders2014}.
\section{Phase estimation with $3/2$-spin cat states}\label{b}
In this section, we shall examine quantum phase estimation using a superposition of $3/2$-spin coherent states, considering scenarios where the phase accumulates through operations involving $e^{i\zeta{S}_{z}}$, $e^{i\zeta{S}_{x}}$ and $e^{i\zeta{S}_{y}}$. To begin, we express the coherent state for a $3/2$-spin system as follows
\begin{align}
	|\theta,\varphi,s\rangle=&\cos^{3}(\frac{\theta}{2})\Big|\frac{3}{2},\frac{-3}{2}\Bigr\rangle+\sqrt{3}e^{-i\varphi}\cos^{2}(\frac{\theta}{2})\sin(\frac{\theta}{2})\Big|\frac{3}{2},\frac{-1}{2}\Bigr\rangle\notag \\
	& +\sqrt{3}e^{-2i\varphi}\cos(\frac{\theta}{2})\sin^{2}(\frac{\theta}{2})\Big|\frac{3}{2},\frac{1}{2}\Bigr\rangle+e^{-3i\varphi}\sin^{3}(\frac{\theta}{2})\Big|\frac{3}{2},\frac{3}{2}\Bigr\rangle.\label{13}
\end{align}
Hence, the superposition of two coherent states of a $3/2$-spin system can be expressed as
\begin{equation}
	|cat,3/2\rangle=N_{c}(|\theta_{1},\varphi_{1},3/2\rangle+|\theta_{2},\varphi_{2},3/2\rangle).
	\label{14}
\end{equation}
Thus, the states given by equation \eqref{14} can also be reformulated in the Dicke basis as
\begin{equation}
	|cat,3/2\rangle=N_{c}\Bigl(A\Big|\frac{3}{2},\frac{-3}{2}\Bigr\rangle+B\Big|\frac{3}{2},\frac{-1}{2}\Bigr\rangle+C\Big|\frac{3}{2},\frac{1}{2}\Bigr\rangle+D\Big|\frac{3}{2},\frac{3}{2}\Bigr\rangle\Bigl),
	\label{15}
\end{equation}
with the normalization factor $N_{c}$ is given by 
\begin{equation}
	N_{c}=\frac{1}{\sqrt{A^{2}+3B\Bar{B}+3C\Bar{C}+D\Bar{D}}},
	\label{16}
\end{equation}
in terms of the coefficients $A$, $B$, $C$ and $D$ given by:
\begin{align}
		&A=\cos^{3}(\frac{\theta_{1}}{2})+\cos^{3}(\frac{\theta_{2}}{2}),\hspace{1cm}B=e^{-i\varphi_{1}}\cos^{2}(\frac{\theta_{1}}{2})\sin(\frac{\theta_{1}}{2})+e^{-i\varphi_{2}}\cos^{2}(\frac{\theta_{2}}{2})\sin(\frac{\theta_{2}}{2}),\notag\\&
		C=e^{-2i\varphi_{1}}\cos(\frac{\theta_{1}}{2})\sin^{2}(\frac{\theta_{1}}{2})+e^{-2i\varphi_{2}}\cos(\frac{\theta_{2}}{2})\sin^{2}(\frac{\theta_{2}}{2}),\hspace{1cm}
		D=e^{-3i\varphi_{1}}\sin^{3}(\frac{\theta_{1}}{2})+e^{-3i\varphi_{2}}\sin^{3}(\frac{\theta_{2}}{2}),
		\label{17}
\end{align}
where the coefficients $\Bar{B}$, $\Bar{C}$ and $\Bar{D}$ are the conjugates of the coefficients $B$, $C$ and $D$, respectively. Our inquiry here pertains to how the spin coherent states would react under the influence of parameters introduced through operators like $e^{i{\zeta}S_{x}}$, $e^{i{\zeta}S_{y}}$ and $e^{i{\zeta}S_{z}}$. While it may initially appear as a mere mathematical curiosity, this question yields profound insights upon closer scrutiny. Notably, within the framework of the Schwinger representation of spin operators, $e^{i{\zeta}S_{x}}$ and $e^{i{\zeta}S_{y}}$ adopt the characteristics of beam-splitting operators, where the transmittance of the beam splitter is determined by the parameter $\zeta$. On the other hand, $e^{i{\zeta}S_{z}}$ assumes the role of a phase shift within an interferometer. Consequently, parameter estimation involving the first pair is concerned with discerning the beam splitter's transmittance, rather than achieving precision of phase shift within an interferometric setup.
\subsection{The phase shift in $3/2$-spin states induced by the operator $S_{z}$}
Now, we will delve into the analysis of the phase sensitivity introduced by $e^{i{\zeta}S_{z}}$, aiming to assess the quantum performance of the state $|Cat,3/2\rangle$ outlined in equation \eqref{15}. Within this context, we encounter the following scenario
\begin{equation}
	|Cat,3/2\rangle_{\zeta}=N_{c}\Bigl(A e^{-i{\zeta}\frac{3}{2}}\Big|\frac{3}{2},\frac{-3}{2}\Bigr\rangle+B e^{-i\frac{\zeta}{2}}\Big|\frac{3}{2},\frac{-1}{2}\Bigr\rangle+C e^{i\frac{\zeta}{2}}\Big|\frac{3}{2},\frac{1}{2}\Bigr\rangle+D e^{i{\zeta}\frac{3}{2}}\Big|\frac{3}{2},\frac{3}{2}\Bigr\rangle\Bigl).
	\label{18}
\end{equation}
As a result, the density matrix characterizing the system is provided by
\begin{equation}
	\rho_{\zeta}=N_{C}^{2}
\begin{pmatrix}
			A^{2} & \sqrt{3} BA e^{i\zeta}  & \sqrt{3} CA e^{2i\zeta}  &
			DA e^{3i\zeta}
			\\
			\sqrt{3}A\Bar{B}e^{-i\zeta}  & 3B\Bar{B} & 3 C\Bar{B}e^{i\zeta} & \sqrt{3}D\Bar{B}e^{2i\zeta}  
			\\
			\sqrt{3}A\Bar{C}e^{-2i\zeta}  & 3B\Bar{C}e^{-i\zeta}  & 3C\Bar{C}& \sqrt{3}D\Bar{C}e^{i\zeta} 
			\\
			A\Bar{D}e^{-3i\zeta}  &\sqrt{3}B\Bar{D}e^{-2i\zeta} & \sqrt{3} C\Bar{D}e^{-i\zeta} & D\Bar{D}
	\end{pmatrix}.
	\label{19}
\end{equation}
We should note that the density matrix $\rho_{\zeta}=|Cat,3/2\rangle_{\zeta~{\zeta}}\langle{Cat,3/2}|$ represents a pure state. In this particular scenario, the QFI pertaining to the operator $S_{z}$ simplifies to
\begin{equation}
	F(\rho_{\zeta},S_{z})=4\Bigr[~_{\zeta} \langle{Cat,3/2}|S_{z}^{2}|{Cat,3/2}\rangle_{\zeta}-\bigr|~_{\zeta}\langle{Cat,3/2}|S_{z}|{Cat,3/2\rangle}_{\zeta}~\bigr|^{2}~\Bigr].
	\label{20}
\end{equation}
By substituting equations \eqref{18} and \eqref{20} into equation \eqref{1}, the expression of the quantum Cramér–Rao bound for the density matrix \eqref{19} takes the form
\begin{equation} 
	\Delta{\zeta}^{S_{z}}=\Bigl\{12~N_{c}^{4}~{\Bigr[{A^{2}(B\Bar{B}+4~C\Bar{C}+3~D\Bar{D})+B\Bar{B}~(3~C\Bar{C}+4~D\Bar{D})+C\Bar{C}~D\Bar{D}}}\Bigr]\Bigl\}^{-\frac{1}{2}}.
	\label{22}
\end{equation}

\begin{figure}[h]
	{{\begin{minipage}[b]{.26\linewidth}
				\centering
				\includegraphics[scale=0.31]{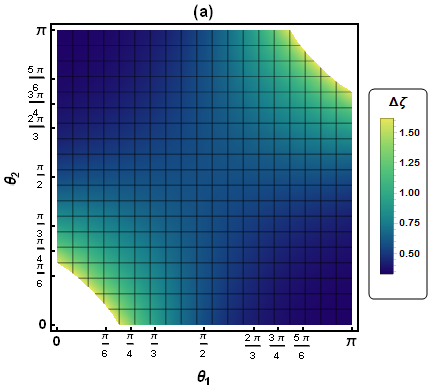} 
			\end{minipage}\hfill
			\begin{minipage}[b]{.26\linewidth}
				\centering
				\includegraphics[scale=0.31]{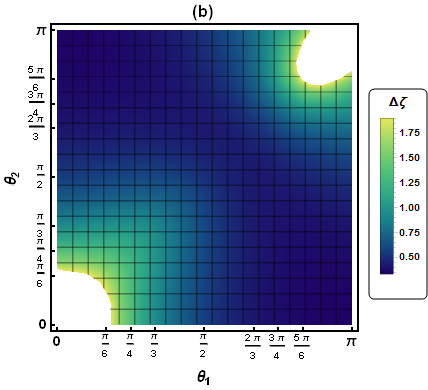} 
		\end{minipage}
		\begin{minipage}[b]{.25\linewidth}
				\centering
				\includegraphics[scale=0.31]{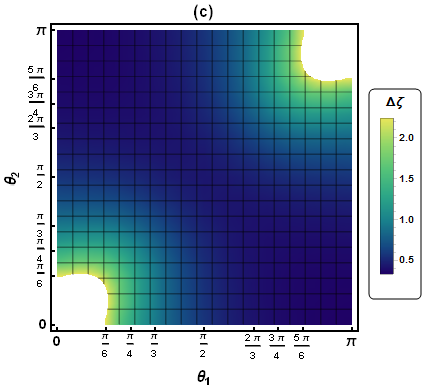} 
			\end{minipage}\hfill
			\begin{minipage}[b]{.22\linewidth}
				\centering
				\includegraphics[scale=0.31]{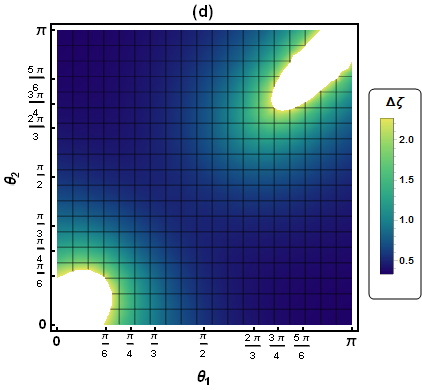} 
	\end{minipage}}}
	\caption{Density of the Cramér–Rao bound $\Delta{\zeta}_{\Phi}^{S_{z}}$ versus $\theta_1$ and $\theta_2$ with $\varphi_{1}=0$. Fig.\textbf{(a)} is obtained for $\varphi_2=0$. Fig.\textbf{(b)} is for  $\varphi_2=\pi/2$. Fig.\textbf{(c)} is for $\varphi_2=4\pi/3$ and Fig.\textbf{(d)} is for $\varphi_2=\pi$.}
	\label{fig1}
\end{figure}

As it can be expected from this result, $\Delta{\zeta}^{S_{z}}$ is symmetric in terms of $\theta_{1}$ and $\theta_{2}$. If we select $\theta_{1}=0$ and $\theta_{2}=\pi$ or vice versa, we achieve the Heisenberg limit, resulting in $\Delta{\zeta}^{S_{z}}=0.333$. From Figure \ref{fig1}($a$), it is clear that $\Delta{\zeta}^{S_{z}}$ diverges when $\theta_{1}=\theta_{2}=0$ and $\theta_{1}=\theta_{2}=\pi$. While the Cramér–Rao bound equals the standard quantum limit for $\theta_{1}=\theta_{2}=\frac{\pi}{2}$, which gives $\Delta{\zeta}^{S_{z}}=0.5773$. Now, considering the case where $\theta_{1}=\theta_{2}\neq{\frac{\pi}{2}}$, the Cramér–Rao bound surpasses the standard quantum limit, indicating that the state does not guarantee maximum precision in the estimation protocol.  Subsequently, it becomes evident that heightened sensitivity is conferred along this axis when $\theta_{2}=\pi-\theta_{1}$.
In Figure \ref{fig1}($b$), we display the behavior of $\Delta{\zeta}_{\Phi}^{S_{z}}$ versus $\theta_{1}$ and $\theta_{2}$ with ${\Phi=\frac{\pi}{2}}$ (see the appendix \ref{AppA}). When $\Phi=\frac{\pi}{2}$, the best performance is achieved for $\theta_{1}=0$ and $\theta_{2}=\pi $ or vice versa. By contrast, the quantum Cramér–Rao bound diverges for $\theta_{1}=\theta_{2}=0$ and $\theta_{1}=\theta_{2}=\pi$. The other interesting situation would be to consider $\Delta{\zeta}$ along the line $\theta_{2}={\pi}-\theta_{1}$. In this case, the value of quantum Cramér–Rao bound is between the Heisenberg limit and the standard quantum limit.\par

In Figure \ref{fig1}($c$), we examine the performance of the density $\Delta{\zeta}$ with ${\Phi=\frac{4\pi}{3}}$ (in this scenario, the analytical expression for the quantum Cramér–Rao bound is given in the appendix \ref{AppA}). The density $\Delta{\zeta}$ exhibits symmetry with respect to $\theta_{1}$ and $\theta_{2}$. Notably, when one of these angles is set to $0$ and the other to $\pi$, the state achieves Heisenberg limit sensitivity, resulting in $\Delta{\zeta} =1/3$. Hence, $\Delta{\zeta}=0.4236$ holds for $\theta_{1}=\theta_{2}$, while it deviates for $\theta_{1} = 0,\pi$. Another intriguing scenario involves considering $\Delta{\zeta}_{\Phi=\frac{4\pi}{3}}^{S_{z}}$ along the line $\theta_{1}=\pi-\theta_{2}$. In this instance, the Cramér–Rao bound assumes values within the range of Heisenberg limit and standard quantum limit (i.e. ${\rm HL}\leq\Delta{\zeta}_{\Phi=\frac{4\pi}{3}}^{S_{z}}\leq{\rm SQL}$).\par

Finally, when we set $\Phi=\varphi_{1}-\varphi_{2}={\pi}$ as depicted in Figure \ref{fig1}($d$) (see the appendix \ref{AppA}), the Cramér–Rao bound reached the Heisenberg limit when $\theta_{1}=0$ and $\theta_{2}=\pi$ or vice versa. Conversely, it diverges for $\theta_{1}=\theta_{2}=0$ and $\theta_{1}=\theta_{2}=\pi$. The other interesting scenario would be to consider $\Delta{\zeta}$ along the line $\theta_{2}=\pi-\theta_{1}$. As a result, the Cramér–Rao bound takes values between the Heisenberg limit and the standard quantum limit. As a consequence, in all these plots, where $\theta_{1}=0$, $\theta_{2}=\pi$ or else $\theta_{1}=\pi$, $\theta_{2}=0$, the phase estimate is optimal. This phenomenon arises due to the fact that, in these instances, the cat state represents a superposition of two antipodal states on the Bloch sphere. Indeed, the superposition of spin coherent states situated antipodally on the Bloch sphere exhibits behaviors akin to NOON states within quantum metrology \cite{Sanders2014}. Thus, the explicit expression of the cat state is reduced to
\begin{equation}
	|Cat,3/2\rangle=N(|\theta_{1},\varphi,3/2\rangle+|\pi-\theta_{1},\varphi,3/2\rangle),
\end{equation} 
where, in this scenario, $\varphi_{1}$ and $\varphi_{2}$ lose their significance at the poles, resulting in phase sensitivity becoming independent of them. Based on the above results, we deduce that the Heisenberg limit is inversely proportional to the $s$-spin according to the relationship $\Delta\zeta_{HL}=1/2s$.\par 

When comparing our results with those of Ref.\cite{Maleki2021} for spin-$1/2$ and spin-$1$, we find nearly identical outcomes. For the case where \(\Phi=0\), the Cramér-Rao bound reaches the Heisenberg limit if one parameter is set to 0 and the other to \(\pi\). This is expected because, in these cases, one of the coherent spin states coincides with the north pole and the other with the south pole of the Bloch sphere. The CRB diverges when all $\theta_{1,2}$ angles are zero or \(\pi\). However, superpositions of these polar states, known as NOON states, achieve the Heisenberg limit. Another interesting situation within this class of cat states arises when \(\theta_2=\pi-\theta_1\). As demonstrated in Fig.3(a) of Ref.\cite{Maleki2021}, the CRB for parameter estimation \(\Delta\xi_{CRB}\) exhibits enhanced sensitivity along this line. For \(\Phi=\pi/2 \), the CRB is optimal when either \( \theta_1=0\) and \(\theta_2 = \pi \), or \(\theta_1 =\pi\) and \( \theta_2=0\) for spins with $s=1/2$, $1$, and $3/2$. When \(\phi=\pi\), \(\Delta \xi_{CRB}\) diverges for \(\theta_1=\theta_2=0\) or \(\pi \), but it is optimal for \(\theta_1=0\) and \(\theta_2=\pi\), or vice versa. In all these scenarios, when \(\theta_1=0\) and \(\theta_2 =\pi\), or alternatively \( \theta_1 = \pi \) and \( \theta_2 = 0 \), phase estimation becomes optimal. This is because, in these cases, the cat state becomes a superposition of the north and south poles of the Bloch sphere, making \(\phi_1 \) and \( \phi_2 \) lose their significance at the poles. Thus, phase sensitivity becomes independent of these phases, resulting in optimal phase estimation.
\subsection{Quantum metrological performance of $3/2$-spin cat states under the influence of generator $S_{x}$}
\begin{figure}[h]
	{{\begin{minipage}[b]{.26\linewidth}
				\centering
				\includegraphics[scale=0.31]{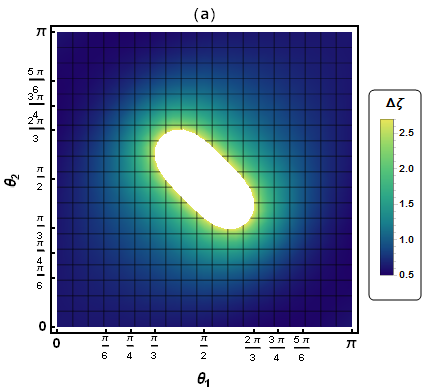} 
			\end{minipage}\hfill
			\begin{minipage}[b]{.26\linewidth}
				\centering
				\includegraphics[scale=0.31]{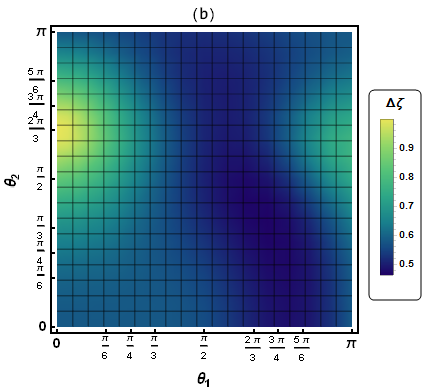} 
		\end{minipage}
	\begin{minipage}[b]{.25\linewidth}
				\centering
				\includegraphics[scale=0.31]{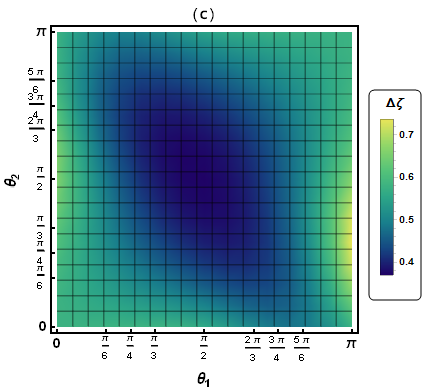} 
			\end{minipage}\hfill
			\begin{minipage}[b]{.22\linewidth}
				\centering
				\includegraphics[scale=0.31]{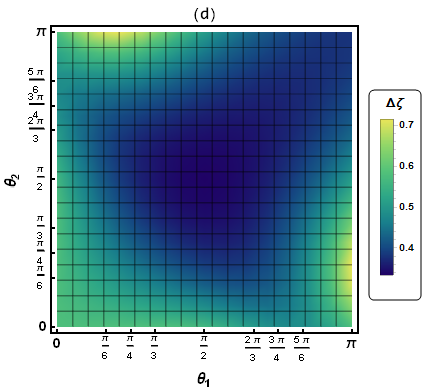} 
	\end{minipage}}}
	\caption{Density of the Cramér–Rao bound $\Delta{\zeta}_{\Phi}^{S_{x}}$ versus $\theta_1$, $\theta_2$ and ${\Phi}=\varphi_{1}-\varphi_{2}$ with $\varphi_1=0$. Fig.\textbf{(a)} is obtained for $\varphi_2=0$. Fig.\textbf{(b)} is for $\varphi_2=\pi/2$. Fig.\textbf{(c)} is for $\varphi_2=3\pi/4$ and Fig.\textbf{(d)} is for $\varphi_2=\pi$.}
	\label{fig2}
\end{figure}

In this second scenario, we examine the metrological performance of $3/2$-spin cat states that accumulates a phase through $e^{i{\zeta}S_{x}} $. The analytical expression for the quantum Cramer–Rao bound is given by
\begin{align}
	\Delta{\zeta}^{S_{x}}=\Bigr[{3N_{c}^{2}{(A^{2}+D\Bar{D}+7B\Bar{B}+7C\Bar{C}+2AC+2A\Bar{C}+2B\Bar{D}+2D\Bar{B})} -{9N_{c}^{4}(AB+A\Bar{B}+2C\Bar{B}+2B\Bar{C}+C\Bar{D}+D\Bar{C})^{2}}}\Bigr]^{\frac{-1}{2}},\label{28}
\end{align}
The general expression of $\Delta{\zeta}^{S_{x}}$ is contingent upon $\varphi_{1}$, $\varphi_{2}$, $\theta_{1}$ and $\theta_{2}$. Consequently, by fixing $\varphi_{1}$ at $0$, we plot $\Delta{\zeta}^{S_{x}}$ in terms of $\theta_{1}$ and $\theta_{2}$ for various values of $\varphi_{2}$, aiming to identify cases of enhanced accuracy. Accordingly, based on the outcomes depicted in Figure (\ref{fig2}), optimal precision is achieved when $\Delta{\zeta}^{S_{x}}$ lies within or equates the range between the standard and Heisenberg quantum limits.

In Figure \ref{fig2}($a$), we have explored the scenario with $\varphi_{2}=0$, resulting in minimal $\Delta{\zeta}^{S_{x}}$ values occurring at $\theta_{1}=\theta_{2}=0,\pi$, with $\Delta{\zeta}^{S_{x}}=0.577$. On the other hand, the highest value of $\Delta{\zeta}^{S_{x}}$ is observed for $\theta_{2}=\pi-\theta_{1}$ within the region of $\frac{\pi}{3}\leq\theta_{2}\leq{\frac{2\pi}{3}}$. In the second case (Fig.~\ref{fig2}(b)) with ${\varphi_2}=\frac{\pi}{2}$, $\Delta{\zeta}^{S_{x}}$ reaches an optimum within the dark blue-shaded region, approximately at the value of $0.4$. For instance, at $\theta_{1}=0$ and $\theta_{1}=\frac{3\pi}{4}$, $\Delta{\zeta}^{S_{x}}=0.479$.\par

The behavior of quantum Cramér–Rao bound as a function of $\theta_{1}$ and $\theta_{2}$, with $\Phi=\frac{3\pi}{4}$ is illustrated in Figure \ref{fig2}($c$). Based on this outcome, it is evident that the estimation is most precise in the region where $\theta_{1}=\theta_{2}=\frac{\pi}{2}$, as the phase measurement error approaches approximately $0.3693$. Furthermore, there exist regions in which $\Delta{\zeta}^{S_{x}}$ provides a better sensitivity: $\frac{\pi}{3}\le{\theta_{1}}\le{\frac{3\pi}{4}}$ and $\frac{\pi}{3}\le{\theta_{2}}\le{\frac{3\pi}{4}}$. For instance, by selecting $\theta_{1}=\frac{2\pi}{3}$ and $\theta_{1}=\frac{\pi}{3}$, we obtain $\Delta{\zeta}^{S_{x}}\simeq {0.3985}$.\par

The identical plot with $\varphi_{2}=\pi$ is depicted in Figure~\ref{fig2}(d). Consequently, $\Delta{\zeta}^{S_{x}}$ is optimized within the region where  $\frac{\pi}{3}\leq\theta_{1}=\theta_{2}\leq\frac{5\pi}{6}$. Another noteworthy scenario in this context involves considering $\theta_{2}=\pi-\theta_{1}$. From Figure \ref{fig2}(d) , we can see that $\Delta{\zeta}^{S_{x}}$ provides a better sensitivity along this line in the region where $\frac{\pi}{3}\leq\theta_{1}=\theta_{2}\leq\frac{2\pi}{3}$. While, if $\theta_{1}=\theta_{2}=\frac{\pi}{2}$, we immediately attain the Heisenberg limit sensitivity (see the appendix \ref{AppB}). 

For the case of estimation with the \( S_x \) generator, if we compare the results with those in paper \cite{Maleki2021} for spin \( s = 1/2 \) and \( \phi = 0 \), the estimation is more accurate and the CRB is optimal when \( \theta_1 = \theta_2 = 0 \) or \( \pi \), and is maximal when \( \theta_1 + \theta_2 = \pi \). Referring back to the spin \( 3/2 \) results, we find similar outcomes, but the behavior of the CRB differs slightly. According to figure 2(a), the area where \( \theta_1 + \theta_2 = \pi \) with \( \theta_{1}\geq \pi/3, \theta_2 \leq 2\pi/3 \) shows that the divergence of the CRB is more sparse compared to the spin \( 1/2 \) case. Comparing \( s = 1/2 \) and \( s = 3/2 \) for \( \phi = \pi/2 \), the CRB is optimal when \( \theta_1 = 0 \) and \( \theta_2 = \pi \), or \( \theta_1 = \theta_2 = 0 \) or \( \pi \), and it diverges when \( \theta_1 = \pi \) and \( \theta_2 = 0 \). For \( s = 3/2 \), the CRB reaches an optimum within the dark blue- shaded region, approximately at a value of 0.4. For instance, at \( \theta_1 = 0 \) and \( \theta_{2}=3\pi/4 \), \( \Delta \zeta_{S_x} = 0.479 \).\par

Finally, for \(\phi_{2}=\pi\), the CRB indicates optimal phase estimation when \( \theta_1 = \theta_2 \). This differs from the case where \(s = 3/2\), where the estimate is accurate when \(\theta_{1}=\theta_{2}= \pi \). The darkest region in Fig.2(d) also represents this scenario \cite{Maleki2021,Maleki22021}.

\subsection{Quantum metrological performance of $3/2$-spin cat states with the parameter-generating operator $S_{y}$}
Finally, we will explore the scenario in which the system accumulates a phase through $e^{i\zeta{S_{y}}}$. Employing a similar approach as described earlier, the Cramer–Rao bound reduces to
\begin{align}
	\Delta{\zeta}^{S_{y}}=\Bigr[{3N_{c}^{2}{(A^{2}+D\Bar{D}+7B\Bar{B}+7C\Bar{C}-2AC-2A\Bar{C}-2B\Bar{D}-2D\Bar{B})}+{9N_{c}^{4}(AB-A\Bar{B}+2C\Bar{B}-2B\Bar{C}-C\Bar{D}+D\Bar{C})^{2}}}\Bigr]^{\frac{-1}{2}},
	\label{37}
\end{align}
and the associated ultimate parameter estimation limit, for the phase difference $\Phi=0$ and $\Phi=\frac{\pi}{2}$, are given by
	\begin{equation}
		\Delta{\zeta}_{\Phi=0}^{S_{y}}=\frac{1}{\sqrt{9-\frac{12}{3-2\cos{\left(\frac{\theta_{1}-\theta_{2}}{2}\right)+\cos{\left(\theta_{1}-\theta_{2}\right)}}}}},\hspace{1.5cm}\Delta{\zeta}_{\Phi=\frac{\pi}{2}}^{S_{y}}=\frac{2\sqrt{2}[-2+\Sigma\cos(\frac{\theta_{1}}{2})\cos(\frac{\theta_{2}}{2})]}{\sqrt{3\Bigr[{\varPi+\Upsilon+\Omega-\Lambda+36}\Bigr]}},\label{38}
	\end{equation}
where
\begin{align}
		&\Sigma=1+\cos(\theta_{1})\Bigl[-2+\cos(\theta_{2})\Bigl]-2\cos(\theta_{2}),\hspace{1cm}
		\varPi=8\cos(\theta_{1})\cos^{2}(\theta_{2})+4\cos(2\theta_{1})\Bigl[1+3\cos^{2}{(\frac{\theta_{1}}{2})}\cos^{3}({\theta_{2}})\Bigl],\notag\\&
		\Upsilon=\cos^{2}{(\frac{\theta_{1}}{2})}\Bigl\{\cos(\theta_{2})\Bigr[13-16\cos(\theta_{1})\cos(2\theta_{2})\Bigr]+7\cos(3\theta_{2})\Bigl\},\hspace{1cm}\Lambda=8\sin^{2}{(\theta_{1})}\cos(2\theta_{2}),\notag\\&
		\Omega=2\cos(\frac{\theta_{1}}{2})\Bigl\{2\cos(\frac{\theta_{2}}{2})\Bigl[-3+\cos(\theta_{1})\Bigl(18-5\cos(\theta_{2})+2\cos(2\theta_{2})+\cos(3\theta_{2})\Bigl)\Bigl]+8\cos(\frac{3\theta_{2}}{2})-2\cos(\frac{7\theta_{2}}{2})\Bigl\},
		\label{40}
\end{align}
and for $\Phi=\frac{\pi}{3}$ and $\Phi=\pi$ are 
\begin{equation}
		\Delta{\zeta}_{\Phi=\frac{\pi}{3}}^{S_{y}}=\frac{32+\Sigma^{'}}{\sqrt{3\Bigr[\varPi^{'}(32+\Sigma^{'})+(-1)^{\frac{2}{3}}12\Upsilon^{'}\Bigr]}}, \hspace{2cm}	\Delta{\zeta}_{\Phi=\pi}^{S_{y}}= \left( 9-\frac{12}{3-2\cos{\left(\frac{\theta_{1}+\theta_{2}}{2}\right)+\cos{\left(\theta_{1}+\theta_{2}\right)}}}\right)^{-\frac{1}{2}},
		\label{41}
	\end{equation} 
where
\begin{equation}
	\begin{split}
		\Sigma^{'}&=9\cos\bigl(\frac{\theta_{1}-3\theta_{2}}{2}\bigl)+9\cos\bigl(\frac{3\theta_{1}-\theta_{2}}{2}\bigl)
		+15\cos\bigl(\frac{\theta_{1}+\theta_{2}}{2}\bigl)-\cos\bigl(\frac{3\theta_{1}+3\theta_{2}}{2}\bigl),\\
		\varPi^{'}&=44+3\cos\bigl(\frac{\theta_{1}-3\theta_{2}}{2}\bigl)+27\cos\bigl(\frac{3\theta_{1}-\theta_{2}}{2}\bigl)-12\cos(2\theta_{2})+5\cos\bigl(\frac{\theta_{1}+\theta_{2}}{2}\bigl)-3\cos\bigl(\frac{3\theta_{1}+\theta_{2}}{2}\bigl),\\
		\Upsilon^{'}&=[1+(-1)^{\frac{1}{3}}]^{2}\sin^{2}\bigl({\frac{\theta_{2}}{2}}\bigl)\Bigr[{\cos\bigl(\frac{\theta_{1}}{2}\bigl)+3\cos\bigl(\frac{3\theta_{1}}{2}\bigl)+3\cos\bigl(\frac{3\theta_{1}}{2}-\theta_{2}\bigl)+8\cos\bigl(\frac{\theta_{2}}{2}\bigl)+\cos\bigl(\frac{\theta_{1}}{2}+\theta_{2}\bigl)}\Bigr]^{2}.
		\label{43}
	\end{split}
\end{equation}
\begin{figure}[h]
	{{\begin{minipage}[b]{.26\linewidth}
				\centering
				\includegraphics[scale=0.31]{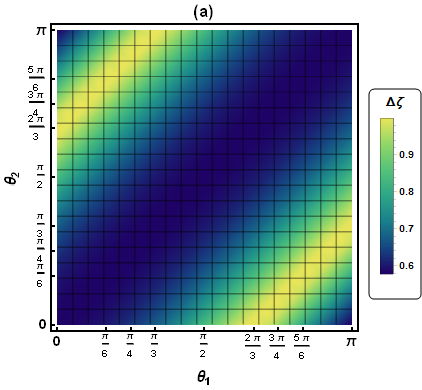} 
			\end{minipage}\hfill
			\begin{minipage}[b]{.26\linewidth}
				\centering
				\includegraphics[scale=0.31]{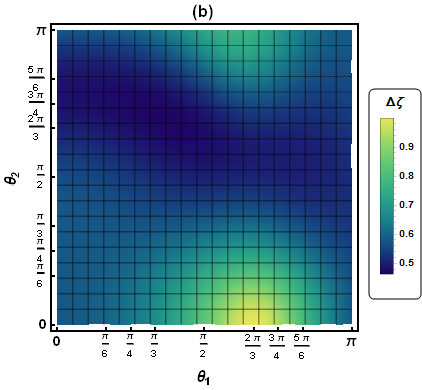} 
			\end{minipage}\begin{minipage}[b]{.25\linewidth}
				\centering
				\includegraphics[scale=0.31]{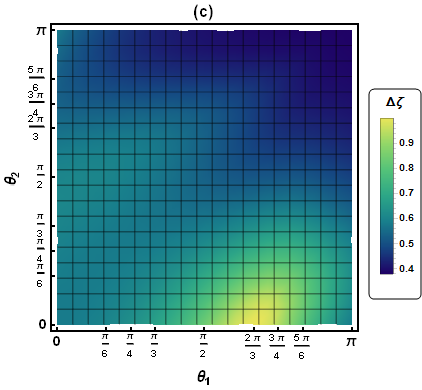} 
			\end{minipage}\hfill
			\begin{minipage}[b]{.22\linewidth}
				\centering
				\includegraphics[scale=0.32]{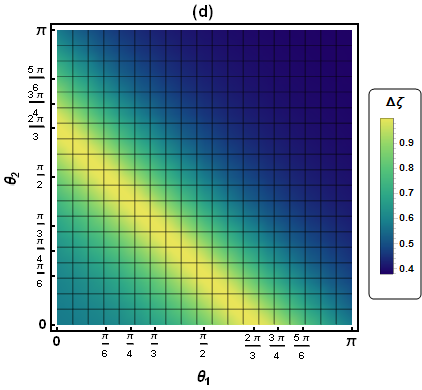} 
	\end{minipage}}}
	\caption{Density of the Cramér–Rao bound $\Delta{\zeta}_{\Phi}^{S_{y}}$ versus $\theta_1$, $\theta_2$ and ${\Phi}=\varphi_1-\varphi_2$ with $\varphi_1=0$. Fig.\textbf{(a)} is obtained for $\varphi_2=0$. Fig.\textbf{(b)} is for $\varphi_2=\pi/2$. Fig.\textbf{(c)} is for $\varphi_2=\pi/3$ and Fig.\textbf{(d)} is for $\varphi_2=\pi$.}
	\label{fig3}
\end{figure} 

From Figure \ref{fig3}($a$), it is evident that the minimum sensitivity occurs when $\theta_{1}=\theta_{2}$, which corresponds to the standard quantum limit. Moreover, the same sensitivity is obtained for cases where $\theta_{1}=0$ and $\theta_{2}=\pi$ or $\theta_{1}=\pi$ and $\theta_{2}=0$. Moving to the second scenario (depicted in Figure \ref{fig3}($b$)), $\Delta{\zeta}^{S_{y}}$ reaches its minimum when $\theta_{1}=0$ and $\theta_{2}=\pi$ or $\theta_{1}=\pi$ and $\theta_{2}=0$, resulting in $\Delta{\zeta}^{S_{y}}=0.471$, a value between the standard quantum limit and the Heisenberg limit. Furthermore, the density Cramer–Rao bound falls below the quantum standard limit within certain regions, such as $0\leq\theta_{1}\leq{\frac{5\pi}{6}}$ and ${\frac{2\pi}{3}}\leq\theta_{2}\leq{\frac{5\pi}{6}}$ simultaneously. As an illustration, for $\theta_{1}=\pi/6$ and $\theta_{2}=5\pi/6$, the density $\Delta{\zeta}^{S_{y}}$ is approximately $0.361$. In the case where $\Phi=\frac{\pi}{3}$ (depicted in Figure \ref{fig3}($c$)), the density Cramer–Rao bound offers improved sensitivity when $\theta_{1}=\theta_{2}=\pi$, as well as in the region surrounding this point. Finally, considering $\Phi=0$ (refer to Figure \ref{fig3}($d$)), we find that $\Delta{\zeta}^{S_{y}}$ equals $0.377964$ when $\theta_{1}=\theta_{2}=\pi$. Clearly, this value approaches the Heisenberg limit $1/3$.
\section{Quantum phase estimations via $s$-spin cat states}\label{x}
\begin{figure}[hbtp]
	{{\begin{minipage}[b]{.33\linewidth}
				\centering
				\includegraphics[scale=0.32]{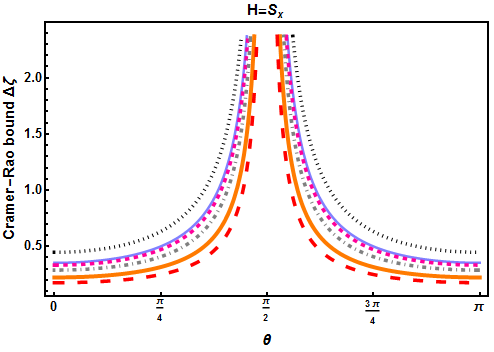} 
			\end{minipage}\hfill
			\begin{minipage}[b]{.33\linewidth}
				\centering
				\includegraphics[scale=0.32]{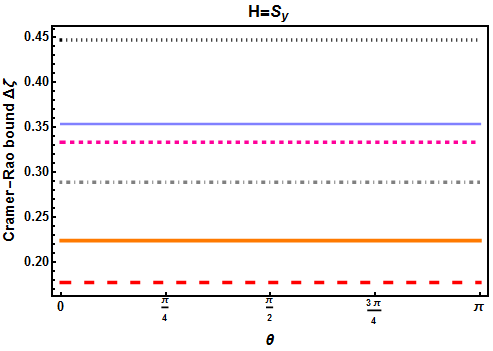}
		\end{minipage}}
		{\begin{minipage}[b]{.33\linewidth}
				\centering
				\includegraphics[scale=0.32]{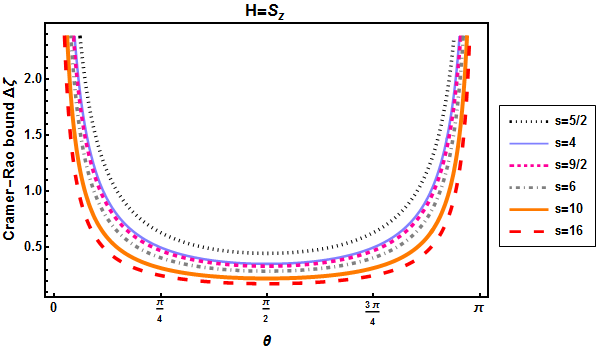} 
			\end{minipage}\hfill}}
	\caption{The Cramér–Rao bound versus $\theta$ for different values of $s$, with ${\Phi}$=$\varphi_1-\varphi_2=0$.} 
	\label{fig4}
\end{figure}

\begin{figure}[hbtp]
	{{\begin{minipage}[b]{.33\linewidth}
				\centering
				\includegraphics[scale=0.32]{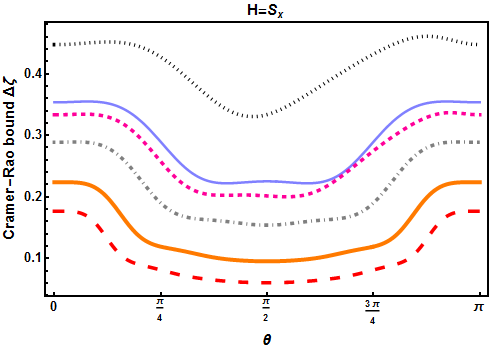} 
			\end{minipage}\hfill
			\begin{minipage}[b]{.33\linewidth}
				\centering
				\includegraphics[scale=0.32]{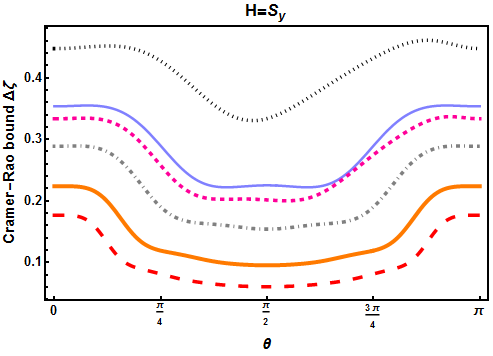} 
		\end{minipage}}
		{\begin{minipage}[b]{.33\linewidth}
				\centering
				\includegraphics[scale=0.32]{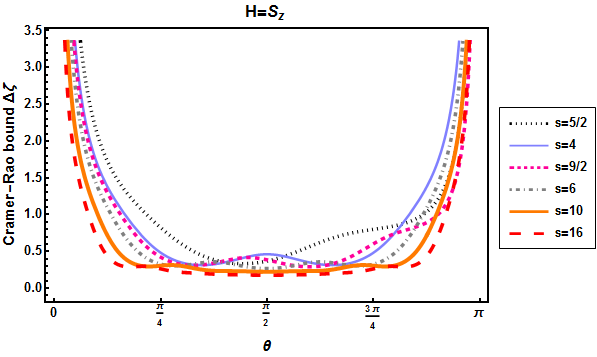} 
			\end{minipage}\hfill}}
	\caption{The Cramér–Rao bound versus $\theta$ for different values of s. ${\Phi}$=$\varphi_1-\varphi_2$=$\pi/2$ in these plots.} 
	\label{fig5}
\end{figure}

\begin{figure}[hbtp]
	{{\begin{minipage}[b]{.33\linewidth}
				\centering
				\includegraphics[scale=0.32]{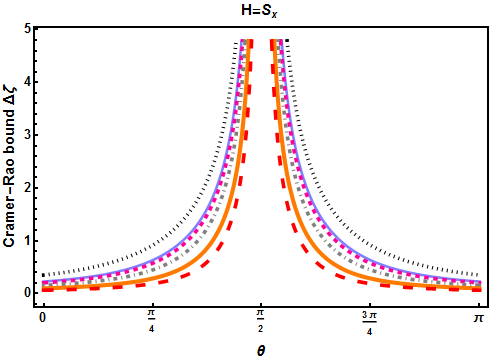} 
			\end{minipage}\hfill
			\begin{minipage}[b]{.33\linewidth}
				\centering
				\includegraphics[scale=0.32]{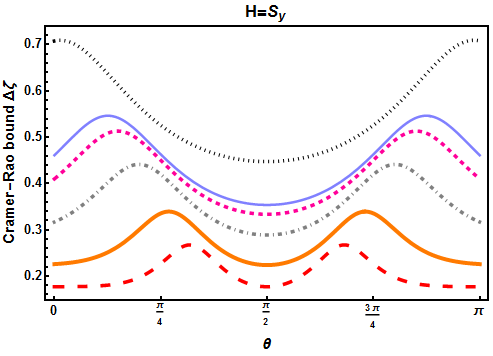} 
		\end{minipage}}
		{\begin{minipage}[b]{.33\linewidth}
				\centering
				\includegraphics[scale=0.32]{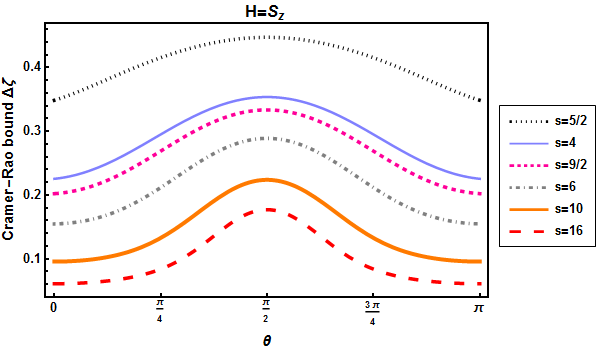} 
			\end{minipage}\hfill}}
	\caption{The Cramér–Rao bound versus $\theta$ for different values of $s$, with ${\Phi}=0$, $\theta=\theta_1$ and $\theta_2=\pi/2$.} 
	\label{fig6}
\end{figure}

\begin{figure}[hbtp]
	{{\begin{minipage}[b]{.33\linewidth}
				\centering
				\includegraphics[scale=0.32]{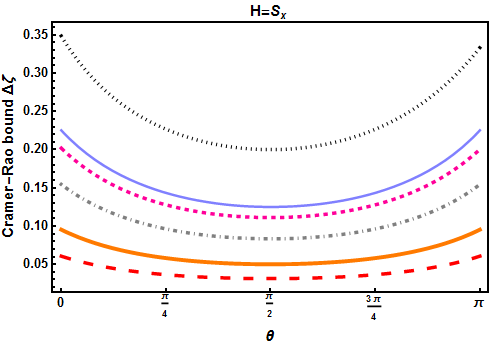} 
			\end{minipage}\hfill
			\begin{minipage}[b]{.33\linewidth}
				\centering
				\includegraphics[scale=0.32]{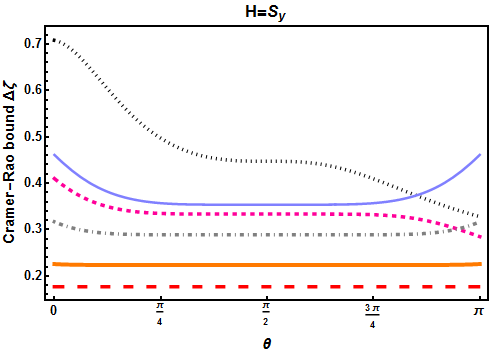} 
		\end{minipage}}
		{\begin{minipage}[b]{.33\linewidth}
				\centering
				\includegraphics[scale=0.32]{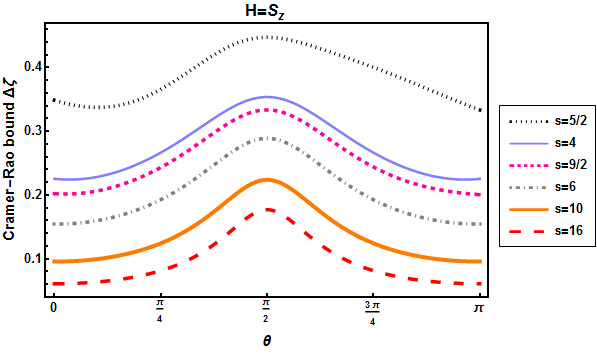} 
			\end{minipage}\hfill}}
	\caption{The Cramér–Rao bound versus $\theta$ for different values of $s$ with ${\Phi}=\pi$, $\theta=\theta_1$ and $\theta_2=\pi/2$.} 
	\label{fig7}
\end{figure}

\begin{figure}[hbtp]
	{{\begin{minipage}[b]{.33\linewidth}
				\centering
				\includegraphics[scale=0.32]{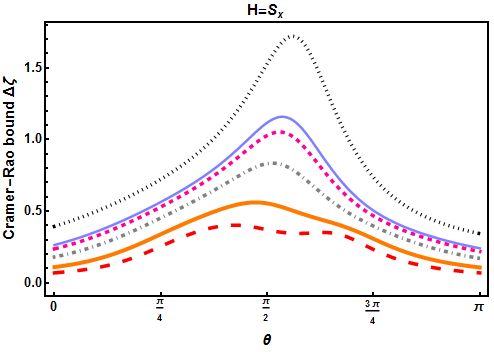} 
			\end{minipage}\hfill
			\begin{minipage}[b]{.33\linewidth}
				\centering
				\includegraphics[scale=0.32]{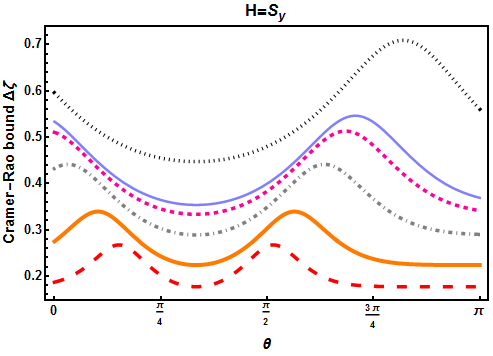} 
		\end{minipage}}
		{\begin{minipage}[b]{.33\linewidth}
				\centering
				\includegraphics[scale=0.32]{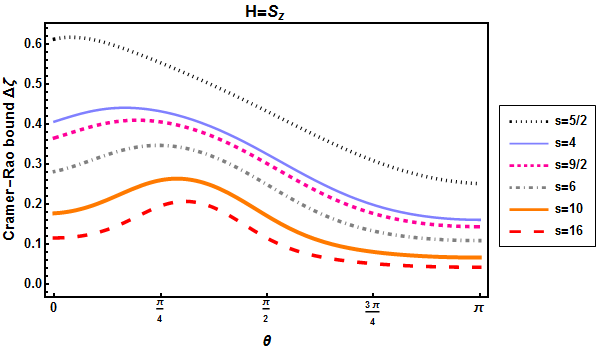} 
			\end{minipage}\hfill}}
	\caption{The Cramér–Rao bound versus $\theta$ for different values of $s$ with ${\Phi}=0$, $\theta=\theta_1$ and $\theta_2=\pi/3$.} 
	\label{fig8}
\end{figure}

Here, we present a comprehensive analytical expression for the quantum Fisher information, facilitating the computation of the Cramér–Rao bound for an unknown parameter within the coherent states superposition of any spin $s$. To investigate the metrological power of a generic superposition of spin coherent states, we derive a general formulation of the QFI for situations where the system accumulates phases via $e^{i\zeta{S}_{x}}$, $e^{i\zeta{S}_{y}}$ and $e^{i\zeta{S}_{z}}$. Notably, we aim to ascertain the precision limitations in each scenario. To achieve this objective, we consider that the states described by \eqref{8} can be alternatively expressed in a more general form as \cite{Wang2000,Slaoui2023,KirdiAppl2023,Durieux2021}
\begin{equation}
	\begin{split}
		|\theta,\varphi,s\rangle&=\left(1+|z|^{2}\right)^{-s}\sum_{m=-s}^{s}{\binom{2s}{s+m}}^{\frac{1}{2}}{z}^{s+m}|s,m\rangle,
	\end{split}
	\label{44}
\end{equation}
where $z=e^{-i\varphi}\tan{(\frac{\theta}{2})}$ and the overlap of two SCSs is 
\begin{equation}
	\begin{split}
		\langle{\theta_{1},\varphi_{1},s} |\theta_{2},\varphi_{2},s\rangle&=\frac{(1+\bar{z_{1}}z_{2})^{2s}}{(1+|z_{1}|^{2})^{s}(1+|z_{2}|^{2})^{s}},
		\label{45}
	\end{split}
\end{equation}
here, $z_{1}=e^{-i\varphi_{1}}\tan{(\frac{\theta_{1}}{2})}$ and $z_{2}=e^{-i\varphi_{2}}\tan{(\frac{\theta_{2}}{2})}$. Therefore, the general form of the superposition of two spin coherent states could be written as
\begin{equation}
	\begin{split}
		|cat,s\rangle&={\cal N}_{c}\left(|\theta_{1},\varphi_{1},s\rangle+|\theta_{2},\varphi_{2},s\rangle\right),\\
		&={\cal N}_{c}\sum_{m=-s}^{s}{\binom{2s}{s+m}}^{\frac{1}{2}}\left[{\frac{{z_{1}}^{s+m}}{(1+|z_{1}|^{2})^{s}}+\frac{{z_{2}}^{s+m}}{(1+|z_{2}|^{2})^{s}}}\right]|s,m\rangle,
		\label{46}
	\end{split}
\end{equation}
in termes of $z_{1}, z_{2} $, and the normalization factor ${\cal N}_{c}$ given by
\begin{equation}
	{\cal N}_{c}=\left[{2+\frac{(1+\bar{z_{1}}z_{2})^{2s}+(1+\bar{z_{2}}z_{1})^{2s}}{(1+|z_{1}|^{2})^{s}(1+|z_{2}|^{2})^{s}}}\right]^{\frac{-1}{2}}.
	\label{47}
\end{equation}
The analytical expressions of the quantum Fisher information for $s$-spin coherent states, in three parameter-generating operators $H=S_{z}$, $H=S_{x}$ and $H=S_{y}$, are proved in appendix \ref{AppC}. In the following, we illustrate the behavior of the quantity $\Delta\zeta$ when the dynamics of the generalized state \eqref{44} is governed by the spin operators $S_{x}$, $S_{y}$, and $S_{z}$. It is noteworthy that in Figures \ref{fig4} and \ref{fig5}, we have taken $\theta=\theta_{1}=\theta_{2}\in[0,\pi]$. For the other figures, specifically Fig.\ref{fig6}, Fig.\ref{fig7} and Fig.\ref{fig8}, we have kept $\theta_{2}$ fixed ($\theta_{2}=\frac{\pi}{2}$) while allowing $\theta_{1}$ to vary within the range $[0,\pi]$ for all three cases.

Figure \ref{fig4} depicts the Cramér–Rao bound plotted against $\theta$ for $\Phi=\phi_{1}-\phi_{2}=0$ and various spin values. When undergoing a unitary evolution along the $x$-direction $(i.e. H=S_{x})$, the Cramér–Rao bound exhibits a minimum within the intervals $[0,\pi/4]$ and $[3\pi/4,\pi]$. If $\theta$ take the values 0 and $\pi$,  the Cramér–Rao bound attains the standard quantum limit. For instance, $\Delta{\zeta}=0.44$ for $s=5/2$, $\Delta{\zeta}=0.35$ for $s=4$ and $\Delta{\zeta}=0.33$ for $s=9/2$ and so forth. Furthermore, the Cramér–Rao bound rises to a maximum value within the ranges $\theta\in[\pi/4,\pi/2[$ and $\theta\in]\pi/2,3\pi/4]$, while it diverges at $\theta=\pi/2$.
When experiencing a unitary evolution along the $y$-direction $(i.e. H = S_{y})$, the Cramér–Rao bound maintains a constant value for each spin, which aligns with the standard quantum limit. Conversely, in the case of a unitary evolution along the $z$-direction $(i.e. H=S_{z})$, maximum precision is achieved for each spin when $\theta=\pi/2$.\par
Based on the outcomes, we observe a consistent pattern in the  Cramér–Rao bound resembling that of a spin $3/2$ across all three directions, with only the Heisenberg limit changing when $s$ changes. Moreover, the figures unmistakably illustrate that the estimation with $S_{y}$ is better, as Cramér–Rao bound remains stable for various values of $s$. Nevertheless, in scenarios where $H=S_x$ and $H=S_z$, $\Delta{\zeta}$ diverges at $\theta=\pi/2$ and $\theta=0,\pi$, respectively.

Figure (\ref{fig5}) presents similar plots, but this time, we consider ${\Phi}=\varphi_{1}-\varphi_{2}=\pi/2$. It is easily seen that the Cramér–Rao bound $\Delta{\zeta}$ behaves along the $x$-direction in a manner akin to its behavior along the $y$-direction. In this scenario, the highest accuracy is attained when $\theta$ falls within the interval $[\pi/4,3\pi/4]$. However, in the case of the $z$-direction, the Cramér–Rao bound is minimum when $\theta=\pi/2$ for $s=16,10,6$. While, the minimum sensitivity happens for $\theta=\pi/3,3\pi/4$ when $s=4$ yielding $\Delta{\zeta}\approx{0.32}$. When considering $s=5/2$, the Cramér–Rao bound is maximal for ${\theta}\in{\rbrack{\pi/3,\pi/2}\lbrack} $ and it gives $\Delta{\zeta}\approx{0.34}$. Finally, in the case of $s=9/2$, $\Delta{\zeta}$ provides a better sensitivity when ${\theta}\in\lbrack{\pi/4,\pi/3}\rbrack$ and ${\theta}\in ]\pi/2,2\pi/3[$, then $\Delta{\zeta}=0.3$ and $\Delta{\zeta}=0.28$ respectively.  

Notably, in Figures (\ref{fig6}), it is striking to observe that the Cramér–Rao bound under the influence of the Hamiltonian $H = S_{x}$ attains its minimum at $\theta = 0,\pi$, with $\Delta\zeta$ values falling between the Heisenberg Limit and the Standard Quantum Limit for the various chosen spin values. However, at $\theta=\pi/2$, $\Delta\zeta$ diverges. For the case of $H=S_{y}$, the Cramér–Rao bound exhibits a noticeable minimum when $\theta=\pi/2$ across the specified spin values. Nonetheless, when $\theta=0,\pi$, $\Delta{\zeta}$ reaches a minimum solely for $s=4,9/2,6,10,16$. By contrast, the Cramér–Rao bound when the Hamiltonian $H =S_{z}$ is  optimal where $\theta=0,\pi$ for differents values of $s$-spin, and $\Delta{\zeta}$ is between SQL and HL. For instance, considering the case of spin $s=4$, we have ${\rm SQL}=0.44\leq{\Delta{\zeta}}\approx{0.35}\leq{\rm HL}=0.2$. This outcome parallels the results for other $s$-spin values.\par

However, the situation is slightly different for ${\Phi}=\pi$ (see Fig.\ref{fig7}). When the input state evolves along the $x$-direction $(H=S_{x})$, the phase estimation becomes minimal in the interval $\theta{\in} \lbrack{\pi/4,3\pi/4}\rbrack $ and especially for $\theta=\pi/2$. In the situation where $H=S_y$, the function $\Delta{\zeta}$ decreases to attain the minimum values at an angle $\theta=\pi$ for half-integer values of spin $s=5/2$ and $s=9/5$, but it’s different for the integer spin $s=4,6,10,16$. On the one hand, we can see from figure, that after an initial decreasing, the Cramér–Rao bound increases for $s=4,6$ and the best estimation of the phase shift parameter is achieved for $\theta=\pi/2$. On the other hand, for $s=10,16$, the value of $\Delta{\zeta}$ remains constant in the interval $[0,\pi]$ and lies between SQL and HL. Finally, in the case where $H=S_z$, the best accuracy is obtained when $\theta=0$ or $\theta=\pi$.


Figures \ref{fig8} depict the Cramér–Rao bound as a function of $\theta$ for various $s$ values with ${\Phi}=0$, $\theta=\theta_{1}$, and $\theta_{2}=\pi/3$. It is evident that, for $H=S_x$, the behavior of the Cramér–Rao bound is similar of that seen in \ref{fig6} and we again observe a minimum value of $\Delta{\zeta}$ at angles $\theta=0,\pi$ for different $s$ values, along with a maximum value at $\theta=0,\pi/2$. When considering a unitary evolution along the y-direction $(H=S_y)$, the Cramér–Rao bound attains its minimum at approximately the midpoint of the interval $[\pi/4,\pi/2]$, as well as at $\theta=\pi$, for various $s$ values. Now, let us turn our attention to the case where $H=S_z$. The optimal estimation is achieved when $\theta=\pi$, and even when $\theta$ is in close proximity to $\pi$, across different $s$ values.\par

\begin{figure}[h]
	{{\begin{minipage}[b]{.16\linewidth}
				\centering
				\includegraphics[scale=0.2]{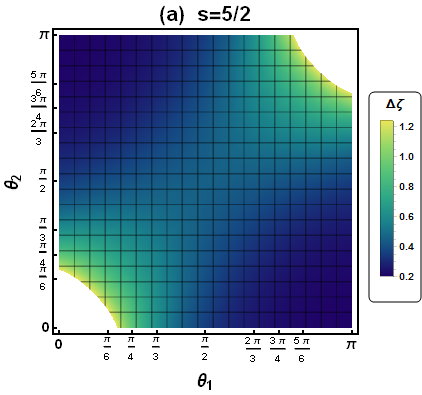} 
			\end{minipage}\hfill
			\begin{minipage}[b]{.16\linewidth}
				\centering
				\includegraphics[scale=0.2]{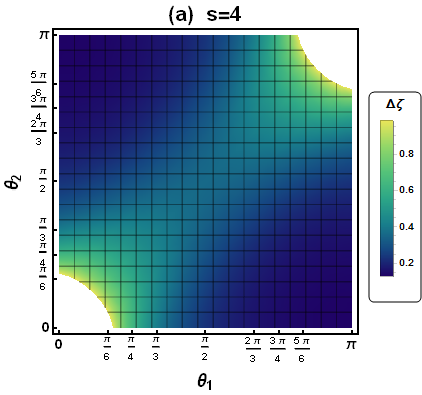} 
			\end{minipage}\hfill
			\begin{minipage}[b]{.16\linewidth}
				\centering
				\includegraphics[scale=0.2]{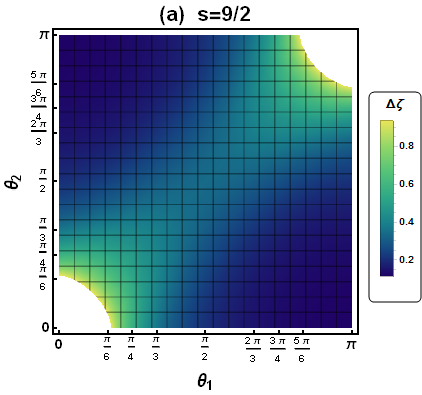}   
			\end{minipage}\hfill
			\begin{minipage}[b]{.16\linewidth}
				\centering
				\includegraphics[scale=0.198]{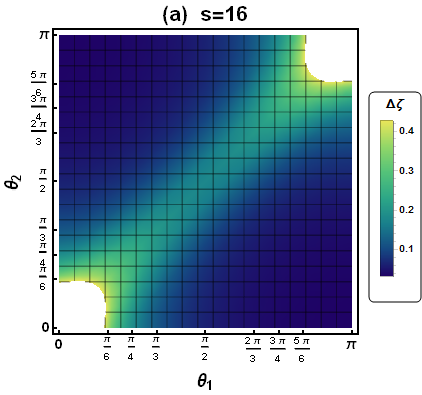}  
	\end{minipage}
\begin{minipage}[b]{.16\linewidth}
	\centering
	\includegraphics[scale=0.2]{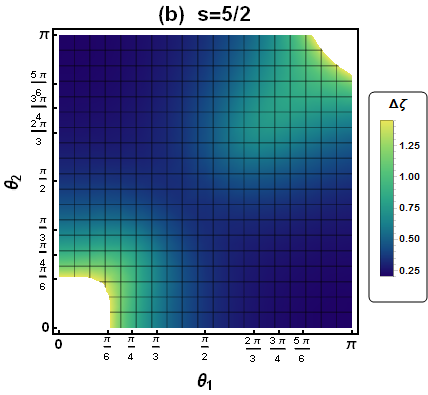}
\end{minipage}\hfill
\begin{minipage}[b]{.16\linewidth}
	\centering
	\includegraphics[scale=0.2]{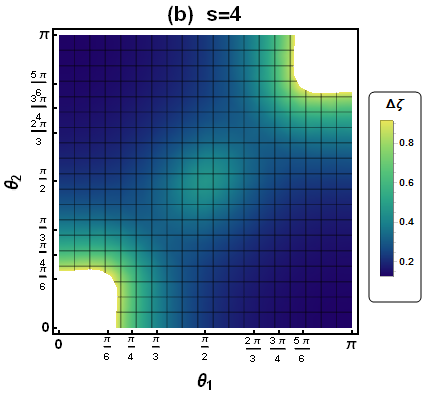}  
\end{minipage}}}
	{{\begin{minipage}[b]{.16\linewidth}
				\centering
				\includegraphics[scale=0.2]{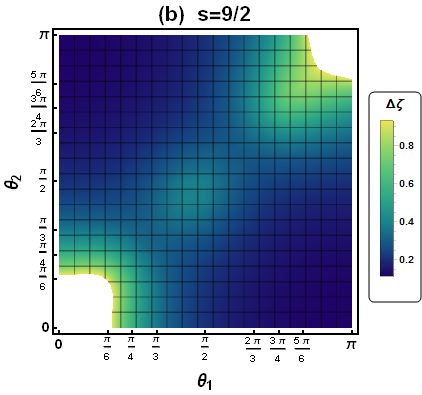}   
			\end{minipage}\hfill
			\begin{minipage}[b]{.16\linewidth}
				\centering
				\includegraphics[scale=0.2]{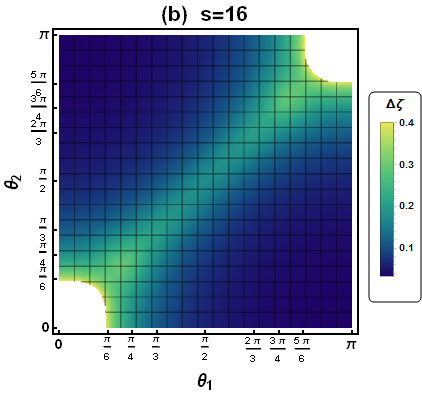}  
			\end{minipage}\hfill 
		\begin{minipage}[b]{.16\linewidth}
				\centering
				\includegraphics[scale=0.2]{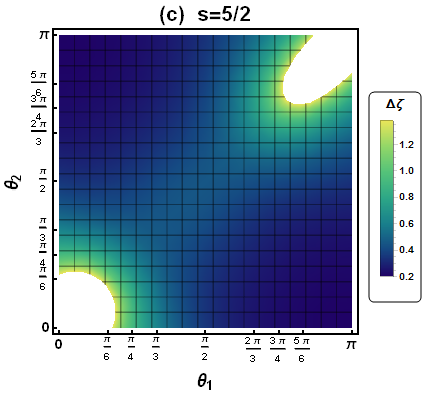} 
			\end{minipage}\hfill
			\begin{minipage}[b]{.16\linewidth}
				\centering
				\includegraphics[scale=0.2]{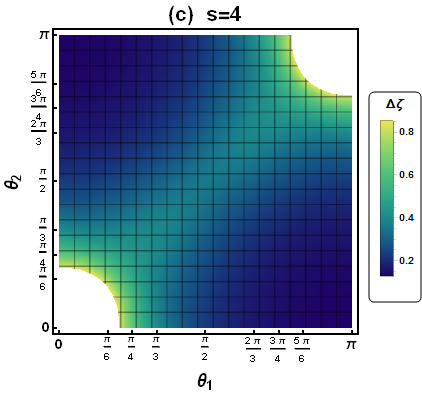} 
			\end{minipage}\hfill
			\begin{minipage}[b]{.16\linewidth}
				\centering
				\includegraphics[scale=0.2]{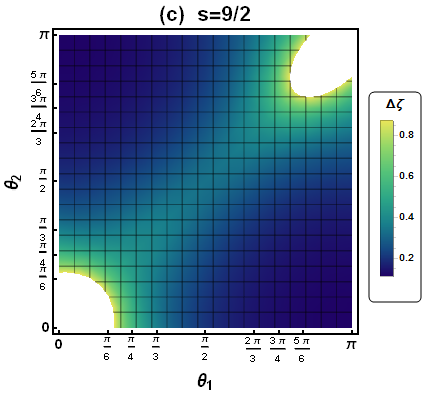} 
			\end{minipage}\hfill
			\begin{minipage}[b]{.16\linewidth}
				\centering
				\includegraphics[scale=0.2]{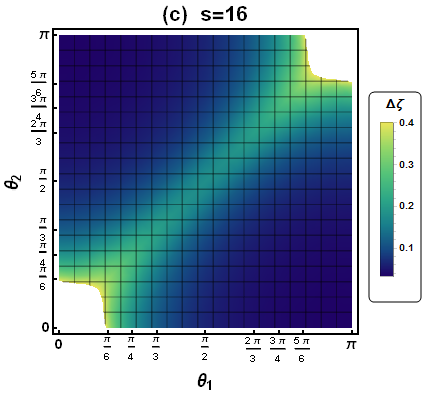} 
	\end{minipage}}}
	\caption{Density of the Cramér–Rao bound versus $\theta_{1}$ and $\theta_{2}$ for different values of $s$ when $H=J_z$. Fig.\textbf{(a)} is for ${\Phi}=0$, Fig.\textbf{(b)} is for  ${\Phi}=\pi/2$, and Fig.\textbf{(c)} is for ${\Phi}=\pi$.}
	\label{fig9}
\end{figure}

\begin{figure}[h]
	{{\begin{minipage}[b]{.16\linewidth}
				\centering
				\includegraphics[scale=0.2]{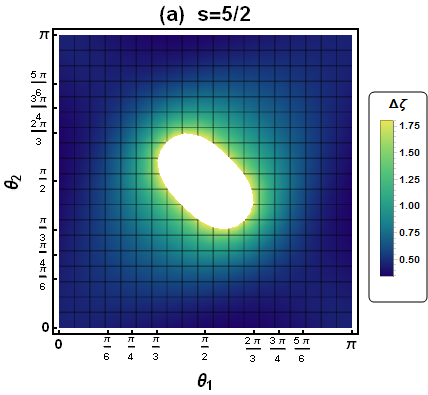} 
			\end{minipage}\hfill
			\begin{minipage}[b]{.16\linewidth}
				\centering
				\includegraphics[scale=0.2]{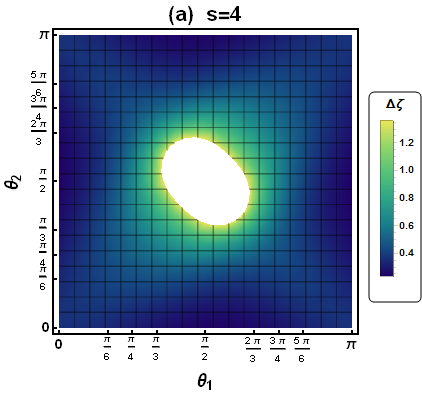}  
			\end{minipage}\hfill
			\begin{minipage}[b]{.16\linewidth}
				\centering
				\includegraphics[scale=0.2]{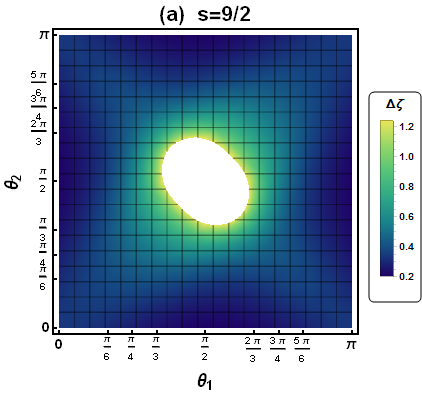}   
			\end{minipage}\hfill
			\begin{minipage}[b]{.16\linewidth}
				\centering
				\includegraphics[scale=0.2]{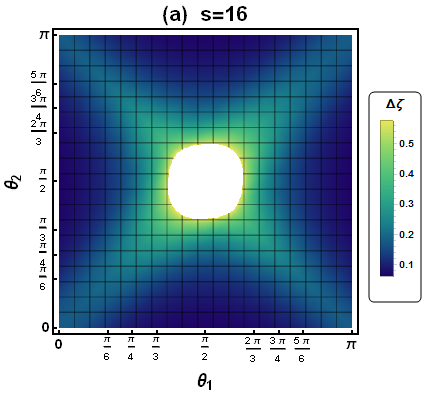}   
	\end{minipage}\hfill
\begin{minipage}[b]{.16\linewidth}
	\centering
	\includegraphics[scale=0.2]{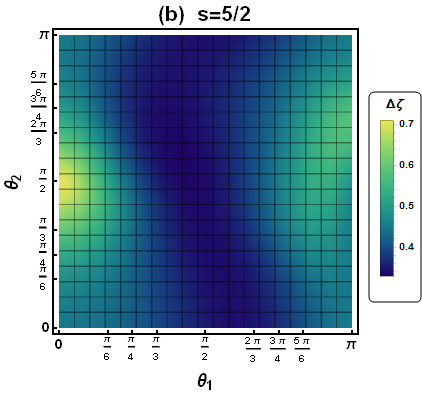}
\end{minipage}\hfill
\begin{minipage}[b]{.16\linewidth}
	\centering
	\includegraphics[scale=0.2]{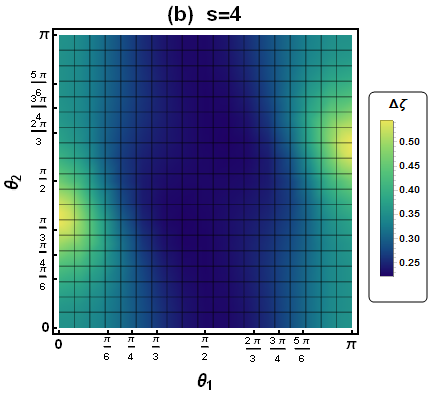}  
\end{minipage}}}
	{{\begin{minipage}[b]{.16\linewidth}
				\centering
				\includegraphics[scale=0.2]{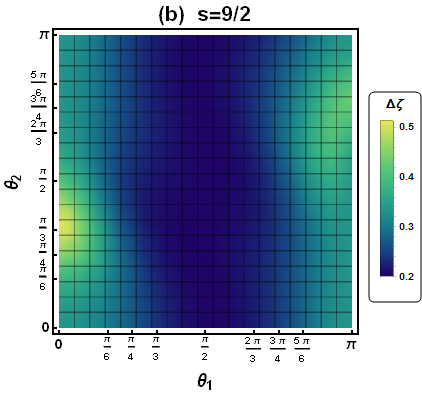}    
			\end{minipage}\hfill
			\begin{minipage}[b]{.16\linewidth}
				\centering
				\includegraphics[scale=0.2]{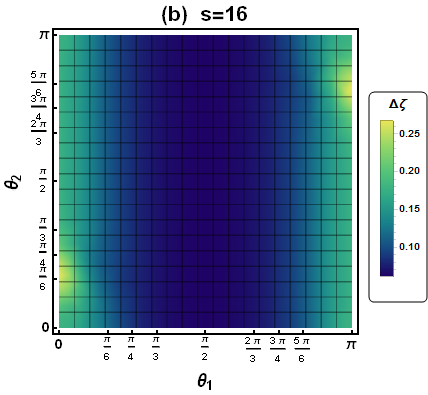}   
			\end{minipage}
		\begin{minipage}[b]{.16\linewidth}
				\centering
				\includegraphics[scale=0.2]{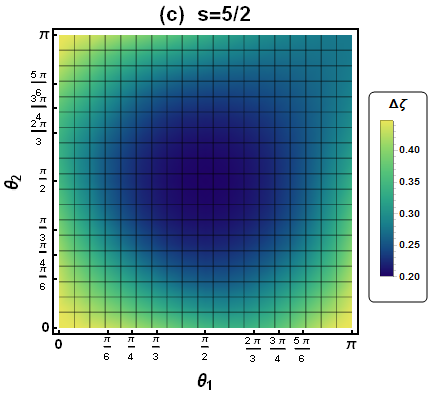} 
			\end{minipage}\hfill
			\begin{minipage}[b]{.16\linewidth}
				\centering
				\includegraphics[scale=0.2]{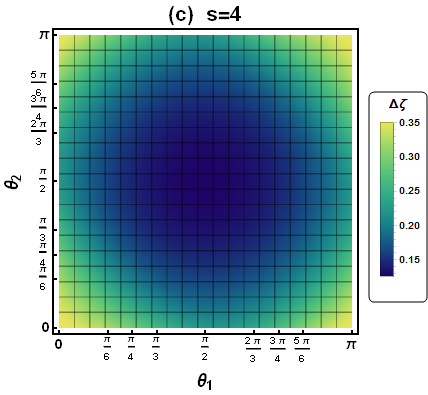} 
			\end{minipage}\hfill
			\begin{minipage}[b]{.16\linewidth}
				\centering
				\includegraphics[scale=0.2]{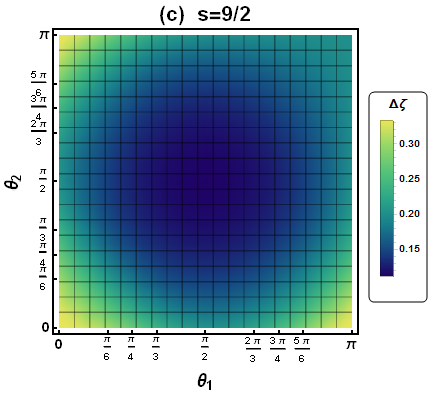} 
			\end{minipage}\hfill
			\begin{minipage}[b]{.16\linewidth}
				\centering
				\includegraphics[scale=0.2]{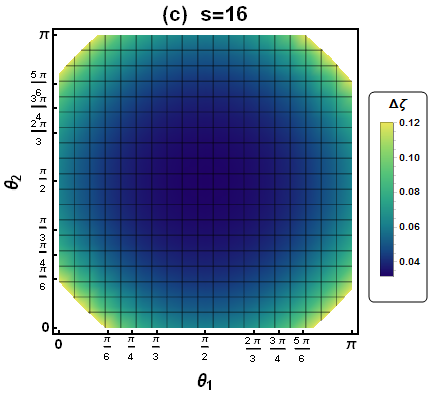} 
	\end{minipage}}}
	\caption{Density of the Cramér–Rao bound versus $\theta_{1}$ and $\theta_{2}$ for different values of $s$ when $H=J_x$. Fig.\textbf{(a)} is for ${\Phi}=0$, Fig.\textbf{(b)} for ${\Phi}$=$\pi/2$, and Fig.\textbf{(c)} is for ${\Phi}$=$\pi$.}
	\label{fig10}
\end{figure}

\begin{figure}[h]
	{{\begin{minipage}[b]{.16\linewidth}
				\centering
				\includegraphics[scale=0.2]{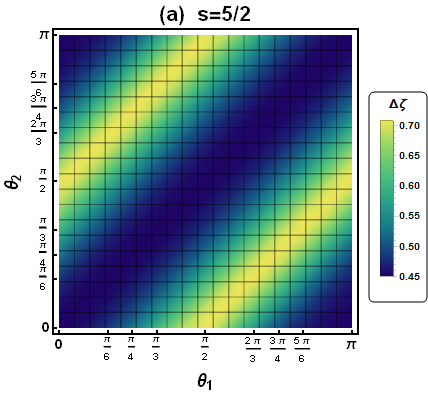} 
			\end{minipage}\hfill
			\begin{minipage}[b]{.16\linewidth}
				\centering
				\includegraphics[scale=0.2]{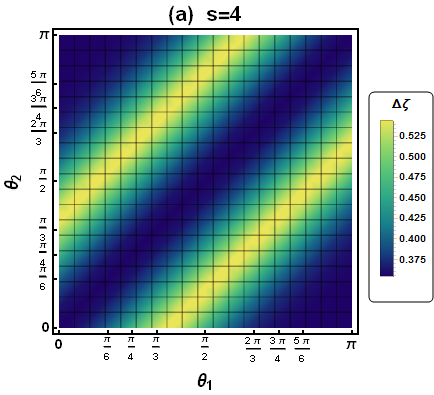}  
			\end{minipage}\hfill
			\begin{minipage}[b]{.16\linewidth}
				\centering
				\includegraphics[scale=0.2]{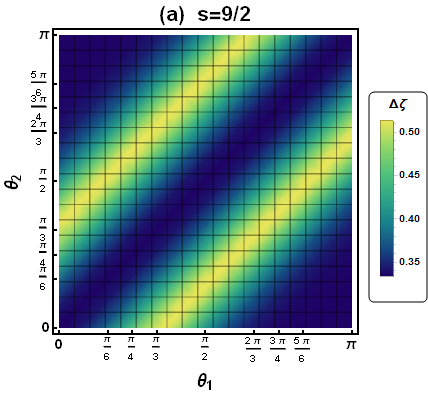}   
			\end{minipage}\hfill
			\begin{minipage}[b]{.16\linewidth}
				\centering
				\includegraphics[scale=0.2]{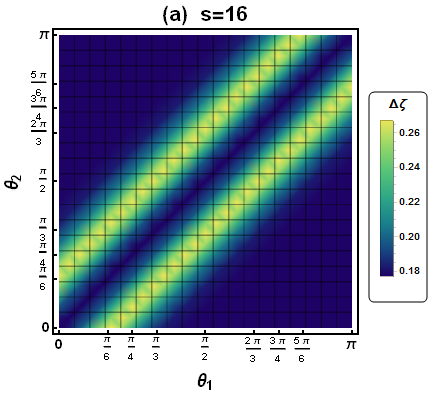}   
	\end{minipage}\hfill
\begin{minipage}[b]{.16\linewidth}
	\centering
	\includegraphics[scale=0.2]{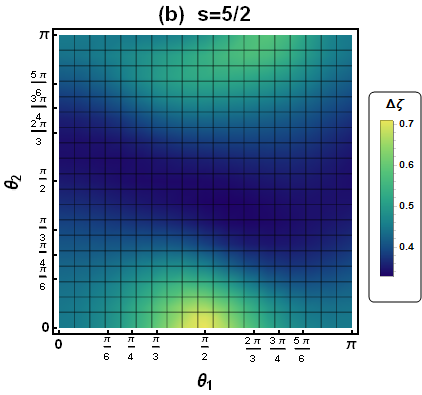}  
\end{minipage}\hfill
\begin{minipage}[b]{.16\linewidth}
	\centering
	\includegraphics[scale=0.2]{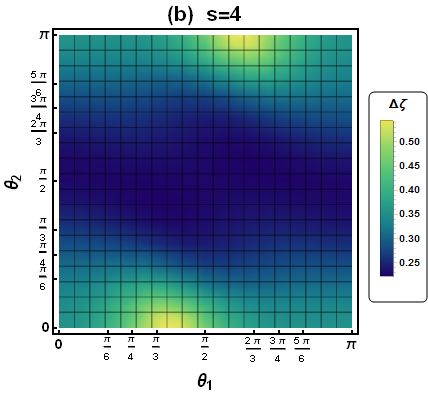}   
\end{minipage}}}
	{{\begin{minipage}[b]{.16\linewidth}
				\centering
				\includegraphics[scale=0.2]{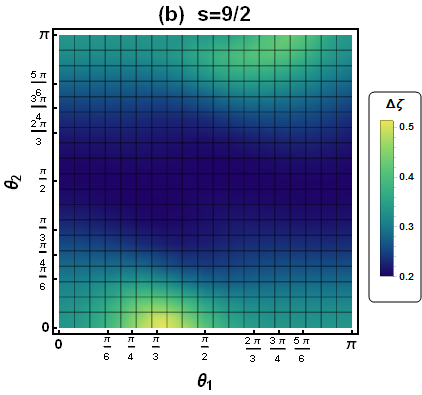}  
			\end{minipage}\hfill
			\begin{minipage}[b]{.16\linewidth}
				\centering
				\includegraphics[scale=0.2]{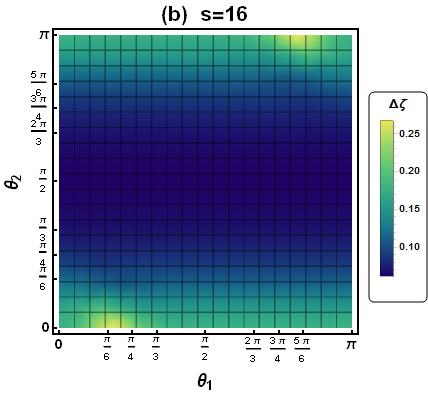}  
			\end{minipage}\hfill
		\begin{minipage}[b]{.16\linewidth}
				\centering
				\includegraphics[scale=0.2]{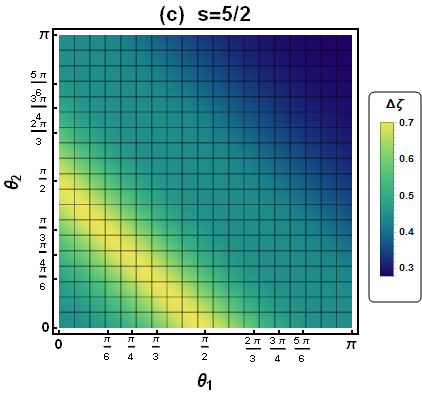} 
			\end{minipage}\hfill
			\begin{minipage}[b]{.16\linewidth}
				\centering
				\includegraphics[scale=0.2]{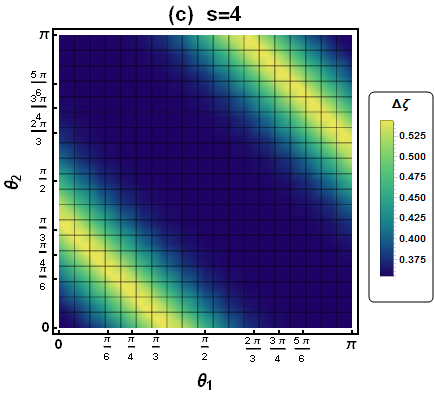} 
			\end{minipage}\hfill
			\begin{minipage}[b]{.16\linewidth}
				\centering
				\includegraphics[scale=0.2]{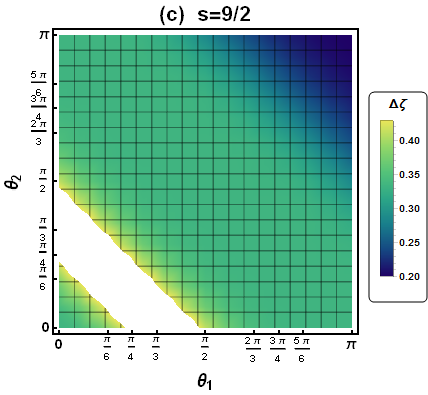} 
			\end{minipage}\hfill
			\begin{minipage}[b]{.16\linewidth}
				\centering
				\includegraphics[scale=0.2]{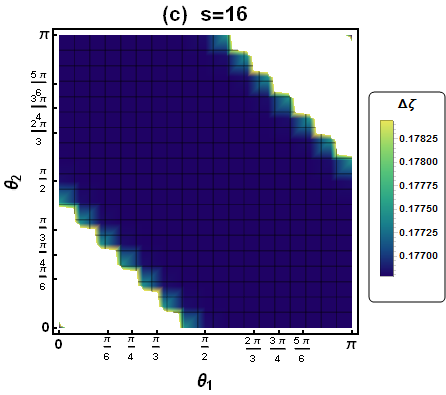} 
	\end{minipage}}}
	\caption{Density of the Cramér–Rao bound versus $\theta_{1}$ and $\theta_{2}$ for different values of $s$ when $H=J_y$. Fig.\textbf{(a)} is for ${\Phi}=0$, Fig.\textbf{(b)} is for  ${\Phi}$=$\pi/2$, and Fig.\textbf{(c)} is for ${\Phi}$=$\pi$.}
	\label{fig11}
\end{figure}

In Figures \eqref{fig9}, \eqref{fig10}, and \eqref{fig11}, we have graphed the CRB as a function of $\theta_{1}$ and $\theta_{2}$ for the x-direction ($H=S_{x}$), y-direction ($H=S_{y}$), and z-direction ($H=S_{Z}$), respectively. These plots consider phase differences $\Phi=\varphi_{1}-\varphi_{2}=0, \pi/2$, and $\Phi=\pi$. The primary aim of these plots is to compare the behavior of the density $\Delta{\zeta}$ for different $s$ values (specifically $s=5/2,4,9/2,16$) with the earlier results obtained for spin $s=3/2$ and to examine how $\Delta{\zeta}$ evolves with increasing $s$. Furthermore, these plots are designed to illustrate scenarios where the Cramér–Rao bound approaches the Heisenberg limit.

To begin with, when comparing the density shapes obtained for different values of $s$ with the previous graphs of $s=3/2$, we observe similar shapes but with varying values and Heisenberg limits as s changes. Clearly, for $H=S_z$ (Fig.\ref{fig9}), the density $\Delta\zeta$ diverges at $\theta_{1}=\theta_{2}=0$ or $\theta_{1}$=$\theta_{2}=\pi$, but is optimal at $\theta_{1}=0$ and $\theta_{2}=\pi$, or vice versa. This is due to the superposition of two antipodal states on the Bloch sphere, resulting in NOON states that exhibit Heisenberg limit precision. Secondly, in the case where $H=S_{x}$ and $\Phi=0$ (Fig.\ref{fig10}($a$)), the Cramér–Rao bound diverges around and into the point $\theta_{1}=\theta_{2}=\frac{\pi}{2}$. Furthermore, the density $\Delta{\zeta}$ is between the quantum standard limit and the Heisenberg limit when ($\theta_{1}=0, \pi$ and $0<\theta_{2}<\pi$) or ($0<\theta_{1}<\pi$ and $\theta_{2}=0, \pi$). For $\Phi=\frac{\pi}{2}$, the Cramér–Rao bound is minimal when ($\frac{\pi}{3}<\theta_{1}<\frac{2\pi}{3}$) for $s=5/2$, and the interval expands as s increases (Fig.\ref{10}($b$)). We set $\Phi=\pi$ in Fig.\ref{10}($c$). In this case, the estimation is more accurate at the point $\theta_{1}=\theta_{2}=\frac{\pi}{2}$.
Finally, when $H = J_{y}$ and $\Phi=0$ (Fig.\ref{fig11}($a$)), the minimum sensitivity  occurs when $\theta_{1}=\theta_{2}$, as well as when ($\theta_{1}=0$ and $\theta_{2}=\pi$) or ($\theta_{1}=\pi$ and $\theta_{2}=0$). In the second case, when $\Phi=\frac{\pi}{2}$ (Fig.\ref{fig11}($b$)), the CRB is minimal when ($\frac{\pi}{3}<\theta_{2}<\frac{2\pi}{3}$) for $s=5/2$, and the interval expands as s increases. In figures \ref{11}($c$), we take $\Phi=\pi$. The density $\Delta{\zeta}$ approaches the Heisenberg limit when $\theta_{1}=\theta_{2}=\pi$. On the other hand, if the spin s is an integer (for example $s=4$ and $s=16$), the density is also minimal in the region where $\theta_{1}=\pi-\theta_{2}$. As $s$ increases in the integer case, the density increases in the regions where $\theta_{1}=\theta_{2}=0, \pi$, and then becomes divergent in the larger values of $s$ (for example $s=16$).\par 

In summary, the minimum value of the Cramér–Rao bound typically falls within the range situated between the Heisenberg limit and the standard quantum limit across various scenarios. It attains the Heisenberg limit under specific conditions contingent upon judiciously chosen parameters $\theta_{1}$, $\theta_{2}$, and the phase difference $\Phi=\varphi_{1}-\varphi_{2}$. Moreover, drawing from preceding findings, it becomes evident that the most precise estimation emerges when $H=S_{z}$, particularly when $\theta_{1}=0$ and $\theta_{2}=\pi$ (or equivalently $\theta_{1}=\pi$ and $\theta_{2}=0$), thereby optimizing the Cramér–Rao bound. This behavior arises from the antipodal nature of the superposed states on the Bloch sphere. Indeed, such superposed spin coherent states, positioned antipodally on the Bloch sphere, exhibit analogous behavior to NOON states within the realm of quantum metrology. However, in instances where $H=S_{x}$ and $H=S_{y}$, the Cramér–Rao bound frequently aligns with the standard quantum limit or resides within the range between the standard quantum limit and the Heisenberg limit. Sometimes, there exist select cases where the Cramér–Rao bound reaches the Heisenberg limit.
\section{Closing remarks} \label{y}
In conclusion, we have thoroughly investigated the metrological performance of superposed $3/2$-spin coherent states and derived an explicit expression for the quantum Cramér–Rao bound specific to this spin configuration. Subsequently, we scrutinized the outcomes for each scenario involving the parameter-generating operator $H$ while considering the Heisenberg limit. To exemplify our study, we derived a general formulation for the quantum Fisher information pertaining to the superposition of $s$-spin coherent states when $H=S_x$, $H=S_y$, $H=S_z$. We presented Cramér–Rao bound plots for each case, showcasing diverse $s$ values. Our investigation reveals that the Heisenberg limit decreases with an increase in the spin number, and this decrease is inversely proportional to $s$, as denoted by the relationship $\Delta\zeta_{\rm HL}=\frac{1}{2s}$. As an example, when $s=1$, ${\rm HL}=0.5$; for $s=3/2$, ${\rm HL}=1/3$; and for $s=4$, ${\rm HL}=0.125$, and so forth. Moreover, our study underscores the profound influence of the chosen parameter-generating operator on quantum metrology and its pivotal role in determining the precision of an unknown parameter. Based on our comprehensive findings, it becomes evident that the operator $S_z$ leads to the most accurate estimations, and with this operator, we reach the Heisenberg limit, resulting in optimal phase estimates. This is because the cat state represents a superposition of two antipodal states on the Bloch sphere ($\theta_1=0$, $\theta_2=\pi$, or $\theta_1=\pi$, $\theta_2=0$). Interestingly enough, $\Delta{\zeta}$ provides HL sensitivity for any choices of the parameters in this case. Based on the results \cite{Maleki22021}, the spin \(s\) and the phase \(\phi\) play a crucial role in the metrological performance of spin coherent states (SCS) in the general form superposition \(N (|\theta, \varphi, s\rangle + |\pi - \theta, \varphi + \phi, s\rangle)\). It's interesting to note that higher photon numbers \(N\) don't necessarily result in higher accuracy of phase estimation. Fine-tuning of the photon number is necessary to achieve the best metrological performance from high \(N\) SCS. An analysis of antipodal SCS on the Bloch sphere indicates that the phase estimation accuracy achievable with such states increases as the distance from the equator of the Bloch sphere increases, reaching the Heisenberg limit for SCS that can be represented as superpositions of the poles of the Bloch sphere, such as NOON states.\par

Indeed, the superposition of spin coherent states located antipodally on the Bloch sphere exhibits behaviour similar to NOON states in quantum metrology. NOON states are entangled quantum states that play a crucial role in enhancing measurement accuracy. In this case, $\varphi_1$ and $\varphi_2$ lose their significance at the poles, rendering the phase sensitivity independent of these phases. As a result, the precision of phase estimation improves as you move away from the equatorial region of the Bloch sphere.\par

To summarize, our article aims to study the behaviour of CRB along the z, y and x directions, and to compare the results obtained. According to the results obtained, the estimation is more accurate with the $S_z$ operator than with $S_x$ and $S_y$. Physically, we think this is because spin coherent states are naturally adapted to this operator; the $S_z$ operator is preferred for quantum phase estimation because of the natural alignment of spin coherent states with this operator. This guarantees accurate phase measurements, as the spin coherent states are eigenstates of $S_z$. On the other hand, $S_x$ and $S_y$ are not naturally matched to these states, which can lead to less accurate phase measurements.\\

{\bf Data Availibility Statement:} No data associated in the manuscript.

\appendix

\section{Analytical expressions of the Cramér–Rao bound for $3/2$-spin states induced by the operator $S_{z}$}\label{AppA}
Upon calculation, the variance $\Delta{\zeta}^{S_{z}}$ is contingent on two parameters $\theta_{1}$ and $\theta_{2}$ as well as the phase difference delineated by $\cos{(\varphi_{1}-\varphi_{2})}$. To understand the characteristics of $\Delta{\zeta}^{S_{z}}$, we will examine distinct scenarios: $\Phi=\varphi_{1}-\varphi_{2}=0$, $\Phi=\frac{\pi}{2}$, $\Phi=\frac{4\pi}{3}$ and $\Phi={\pi}$. In these contexts, we will establish the boundaries for phase estimation through $e^{i{\zeta}S_{z}}$ to assess the efficacy of states conforming to the form outlined in equation \eqref{18}. Clearly, for $\Phi=\varphi_{1}-\varphi_{2}=0$, the density of the Cramér–Rao bound \eqref{22} manifests as
\begin{equation} 
	\Delta{\zeta}_{\Phi=0}^{S_{z}}=\Bigl\{12N_{\Phi=0}^{4}{\Bigr[{A^{2}(\xi+4\kappa+3\vartheta)+\xi(3\kappa+4\vartheta)+\kappa\vartheta}}\Bigr]\Bigl\}^{-\frac{1}{2}},
\end{equation}
where the normalization factor and the associated coefficients are
\begin{equation}
	\begin{split}
		N_{\Phi=0}&=\Bigl\{{\frac{1}{2}}\Bigr[{4+3\cos{(\frac{{\theta_{1}}-{\theta_{2}}}{2})}+\cos{(\frac{3{\theta_{1}}-3{\theta_{2}}}{2})}}\Bigr]\Bigl\}^{-1/2},\hspace{1cm}
		\xi=\Bigr[{\cos^{2}(\frac{\theta_{1}}{2})\sin(\frac{\theta_{1}}{2})+\cos^{2}(\frac{\theta_{2}}{2})\sin(\frac{\theta_{2}}{2})}\Bigr]^{2},\\
		\kappa&=\Bigr[{\cos(\frac{\theta_{1}}{2})\sin^{2}(\frac{\theta_{1}}{2})+\cos(\frac{\theta_{2}}{2})\sin^{2}(\frac{\theta_{2}}{2})}\Bigr]^{2},\hspace{2cm}
		\vartheta=\Bigr[\sin^{3}{(\frac{\theta_{1}}{2})}+\sin^{3}{(\frac{\theta_{2}}{2})}\Bigr]^{2}.
	\end{split}
	\label{23}
\end{equation}
In Figure \ref{fig1}($b$), we display the behavior of $\Delta{\zeta}_{\Phi}^{S_{z}}$ versus $\theta_{1}$ and $\theta_{2}$ with ${\Phi=\frac{\pi}{2}}$. In this case, the quantum Cramér–Rao bound \eqref{22} reduces to
\begin{equation} 
	\Delta{\zeta}_{\Phi=\frac{\pi}{2}}^{S_{z}}=\Bigl\{12N_{\Phi=\frac{\pi}{2}}^{4}{\Bigr[{A^{2}(\xi^{'}+4\kappa^{'}+3\vartheta^{'})+\xi^{'}(3\kappa^{'}+4\vartheta^{'})+\kappa^{'}\vartheta^{'}}}\Bigr]\Bigl\}^{-\frac{1}{2}},
\end{equation}
with
\begin{align}
	&N_{\Phi=\frac{\pi}{2}}=2\Bigl\{8+\cos{(\frac{\theta_{1}}{2})}\Bigr[6\cos(\theta_{1})\cos{(\frac{\theta_{2}}{2})}-2\cos(\frac{3\theta_{2}}{2})(-2+\cos(\theta_{1}))\Bigr]\Bigl\}^{\frac{-1}{2}},\hspace{1cm}\vartheta^{'}=\sin^{6}{(\frac{\theta_{1}}{2})}+\sin^{6}{(\frac{\theta_{2}}{2})},\notag\\& 
	\xi^{'}=\Bigr[{\cos^{4}{(\frac{\theta_{1}}{2})}\sin^{2}{(\frac{\theta_{1}}{2})}+\cos^{4}{(\frac{\theta_{2}}{2})}\sin^{2}{(\frac{\theta_{2}}{2})}}\Bigr],\hspace{1cm}\kappa^{'}=\Bigr[{\cos{(\frac{\theta_{1}}{2})}\sin^{2}{(\frac{\theta_{1}}{2})}-\cos{(\frac{\theta_{2}}{2})}\sin^{2}{(\frac{\theta_{2}}{2})}}\Bigr]^{2},
	\label{24}
\end{align}
In Figure \ref{fig1}($c$), we examine the performance of the density $\Delta{\zeta}$ with ${\Phi=\frac{4\pi}{3}}$. In this scenario, we obtain
\begin{equation} 
	\Delta{\zeta}_{\Phi=\frac{4\pi}{3}}^{S_{z}}=\Bigl\{12N_{\Phi=\frac{4\pi}{3}}^{4}{\Bigr[{A^{2}(\xi^{''}+4\kappa^{''}+3\vartheta^{''})+\xi^{''}(3\kappa^{''}+4\vartheta^{''})+\kappa^{''}\vartheta^{''}}}\Bigr]\Bigl\}^{-\frac{1}{2}},
\end{equation}
and
\begin{equation}
	\begin{split}
		N_{\Phi=\frac{4\pi}{3}}&=4\Bigl[32+15\cos(\frac{\theta_{1}-\theta_{2}}{2})-\cos(\frac{3\theta_{1}-3\theta_{2}}{2})+9\cos(\frac{3\theta_{1}+\theta_{2}}{2})+9\cos(\frac{\theta_{1}+3\theta_{2}}{2})\Bigl]^{\frac{-1}{2}},\\ 
		\xi^{''}&=\frac{1}{16}\Bigr[{\csc^{2}(\frac{\theta_{1}}{2})\sin^{4}(\theta_{1})-\csc(\frac{\theta_{1}}{2})\csc(\frac{\theta_{2}}{2})\sin^{2}(\theta_{1})\sin^{2}(\theta_{2})+\csc^{2}(\frac{\theta_{2}}{2})\sin^{4}(\theta_{2})}\Bigr],\\
		\kappa^{''}&=\frac{1}{4}\Bigr[{\sin^{2}(\frac{\theta_{1}}{2})\sin^{2}(\theta_{1})-\sin(\frac{\theta_{1}}{2})\sin(\frac{\theta_{2}}{2})\sin(\theta_{1})\sin(\theta_{2})+\sin^{2}(\frac{\theta_{2}}{2})\sin^{2}(\theta_{2})}\Bigr],\hspace{0.5cm}
		\vartheta^{''}=\Bigr[\sin^{3}{(\frac{\theta_{1}}{2})}+\sin^{3}{(\frac{\theta_{2}}{2})}\Bigr]^{2}.
	\end{split}
	\label{25}
\end{equation}
Finally, when we set $\Phi=\varphi_{1}-\varphi_{2}={\pi}$ as depicted in Figure \ref{fig1}($d$), we can derive the quantum Cramér–Rao bound as follows
\begin{equation} 
	\Delta{\zeta}_{\Phi=\pi}^{S_{z}}=\Bigl\{12N_{\Phi=\pi}^{4}{\Bigr[{A^{2}(\xi^{'''}+4\kappa^{'''}+3\vartheta^{'''})+\xi^{'''}(3\kappa^{'''}+4\vartheta^{'''})+\kappa^{'''}\vartheta^{'''}}}\Bigr]\Bigl\}^{-\frac{1}{2}},
\end{equation}
with
\begin{align}
	&{N_{\Phi=\pi}=\Bigl\{{{\frac{1}{2}}\Bigr[4+3\cos{(\frac{{\theta_{1}}+{\theta_{2}}}{2})}+\cos{(\frac{3{\theta_{1}}+3{\theta_{2}}}{2})}}}\Bigr]\Bigl\}^{-1/2}, \hspace{1cm}
	\xi^{'''}=\Bigr[{\cos^{2}(\frac{\theta_{1}}{2})\sin(\frac{\theta_{1}}{2})-\cos^{2}(\frac{\theta_{2}}{2})\sin(\frac{\theta_{2}}{2})}\Bigr]^{2},\notag\\& 
	\kappa^{'''}=\Bigr[{\cos(\frac{\theta_{1}}{2})\sin^{2}(\frac{\theta_{1}}{2})+\cos(\frac{\theta_{2}}{2})\sin^{2}(\frac{\theta_{2}}{2})}\Bigr]^{2},\hspace{1cm}
	\vartheta^{'''}=\Bigl[\sin^{3}{(\frac{\theta_{1}}{2})}-\sin^{3}{(\frac{\theta_{2}}{2})}\Bigl]^{2}.
	\label{26}
\end{align}
\section{Analytical expressions of the Cramér–Rao bound for $3/2$-spin cat states under the influence of generator $S_{x}$}\label{AppB}
In the first case where $\Phi=0$ (Fig.~\ref{fig2}(a)), the Cramer–Rao bound is simplified to
\begin{equation}
	\Delta{\zeta}_{\Phi=0}^{S_{x}}=\frac{\sqrt{2}\lbrack{3-2\cos(\frac{\theta_{1}-\theta_{2}}{2})+\cos(\theta_{1}-\theta_{2})}\rbrack}{\sqrt{3}\lbrack{9\eta-2\delta-18\gamma-36\varpi+26\upsilon+14\lambda+\varepsilon+29}\rbrack^{\frac{1}{2}}},
	\label{29}
\end{equation}
where
\begin{align}
	&\eta=\cos(2\theta_{1})+\cos(2\theta_{2}),\hspace{1cm}\delta=\cos\bigl(\frac{5\theta_{1}-\theta_{2}}{2}\bigl)+\cos\bigl(\frac{\theta_{1}-5\theta_{2}}{2}\bigl)+2\cos\bigl(\frac{3\theta_{1}-3\theta_{2}}{2}\bigl),\hspace{1cm}\lambda=\cos(\theta_{1}-\theta_{2}),\notag\\&
	\gamma=\cos\bigl(\frac{3\theta_{1}+\theta_{2}}{2}\bigl)+\cos\bigl(\frac{\theta_{1}+3\theta_{2}}{2}\bigl),\hspace{1cm}\varpi=\cos\bigl(\frac{1}{2}(\theta_{1}-\theta_{2})\bigl),\hspace{1cm}\upsilon=\cos(\theta_{1}+\theta_{2}),\hspace{1cm}
	\varepsilon=\cos(2(\theta_{1}-\theta_{2})).
	\label{30}
\end{align}
In the second case (Fig.~\ref{fig2}(b)) with ${\varphi_2}=\frac{\pi}{2}$, The formulation of the obtained Cramer–Rao bound is as follows:
\begin{equation}
	\Delta{\zeta}_{\Phi=\frac{\pi}{2}}^{S_{x}}=\frac{4(-2+\eta^{'}\cos(\frac{\theta_{1}}{2})\cos(\frac{\theta_{2}}{2}))}{\sqrt{3}\lbrack{\delta^{'}+\lambda^{'}+\gamma^{'}+\varpi^{'}-\upsilon^{'}+\varepsilon^{'}+72}\rbrack^{\frac{1}{2}}},
	\label{31}
\end{equation}
in terms of the quantities
\begin{align}
	&\eta^{'}=1+\cos(\theta_{1})[-2+\cos(\theta_{2})]-2\cos(\theta_{2}),\hspace{1cm}
	\delta^{'}=9\cos(\theta_{1})-8\cos(2\theta_{1})+3\cos(3\theta_{1}),\notag\\&
	\lambda^{'}=8\cos(\theta_{1})\cos(\frac{\theta_{1}}{2})\cos(\frac{\theta_{2}}{2})\Bigl[{5+6\cos(\theta_{1})-3\cos(2\theta_{1})}\Bigl],\hspace{1cm}\gamma^{'}=\cos(\theta_{1})\cos(\theta_{2})\Bigl[{9+16\cos(\theta_{1})+\cos(2\theta_{1})}\Bigl],\notag\\
	&\varpi^{'}=4\cos(\frac{\theta_{1}}{2})\cos(\frac{3\theta_{2}}{2})\Bigl[{18-5\cos(\theta_{1})+2\cos(2\theta_{1})+\cos(3\theta_{1})}\Bigl],\notag\\&\upsilon^{'}=2\cos(\theta_{1})\cos(2\theta_{2})\Bigl[{-3 -8\cos(\theta_{1})+\cos(2\theta_{1})}\Bigl],\hspace{1cm}\varepsilon^{'}=6\cos^{3}(\theta_{1})\cos(3\theta_{2}).
	\label{32}
\end{align}
For $\Phi=\frac{3\pi}{4}$ (Figure \ref{fig2}($c$)), one can readily derive the subsequent expression
\begin{equation} 
	\Delta{\zeta}_{\Phi=\frac{3\pi}{4}}^{S_{x}}=\frac{\eta^{''}}{2\sqrt{3}\Bigr[-\frac{3}{4}(\delta^{''}+\lambda^{''})^{2}-{\frac{1}{2}\gamma^{''}\varpi^{''}(\sqrt{2}+3\sqrt{2}\upsilon^{''}+\varepsilon^{''}+\omega)\Bigr]^{\frac{1}{2}}}},
	\label{33}
\end{equation}
where
\begin{align}
	&\eta^{''}=4+4\cos^{3}(\frac{\theta_{1}}{2})\cos^{3}(\frac{\theta_{2}}{2})-\sqrt{2}\sin(\frac{\theta_{1}}{2})\sin(\frac{\theta_{2}}{2})\Bigr[1+2\cos({\theta_{1}})+\cos({\theta_{2}})\bigr(2+\cos({\theta_{1}})\bigr)\Bigr],\notag\\&
	\delta^{''}=\sqrt{2}\Bigl\{\cos(\frac{\theta_{1}}{2})\sin(\frac{\theta_{2}}{2})\Bigr[-2\cos{(\theta_{2})}+\cos{(\theta_{1})}\cos{(\theta_{2})}-1\Bigr]-\sin({\theta_{2}})\Bigl\},\hspace{1cm}\lambda^{''}=2\sin{(\theta_{1})}\Bigr[1+\cos(\frac{\theta_{1}}{2})\cos^{3}(\frac{\theta_{2}}{2})\Bigr],\notag\\&
	\gamma^{''}=\sqrt{2}+2\csc^{3}{(\frac{\theta_{1}}{2})}\csc^{3}{(\frac{\theta_{2}}{2})}+\cot^{2}{(\frac{\theta_{1}}{2})}\cot^{2}{(\frac{\theta_{2}}{2})}\Bigr[{-3\sqrt{2}+2\cot{(\frac{\theta_{1}}{2})}\cot{(\frac{\theta_{2}}{2})}}\Bigr],\hspace{1cm}\varpi^{''}=\sin^{6}(\frac{\theta_{1}}{2})\sin^{4}(\frac{\theta_{2}}{2}),\notag\\&
	\upsilon^{''}=\csc^{2}(\frac{\theta_{1}}{2})\cos(\theta_{2})\Bigr[2+\cos({\theta_{1}})\Bigr],\hspace{1cm}\varepsilon^{''}=\csc^{3}{(\frac{\theta_{1}}{2})}\csc{(\frac{\theta_{2}}{2})}\Bigr[{-7+2\cos({2\theta_{1}})+\cos({2\theta_{2}})}\Bigr],\notag\\&
	\omega=\cot{(\frac{\theta_{1}}{2})}\Bigr[{5\sqrt{2}}\cot{(\frac{\theta_{1}}{2})}-8\cot{(\frac{\theta_{2}}{2})}-4\cos^{2}{(\frac{\theta_{2}}{2})}\cot^{2}{(\frac{\theta_{1}}{2})}\cot{(\frac{\theta_{2}}{2})}+4\sin{(\theta_{2})}\Bigr].
	\label{34}
\end{align}
For $\varphi_{2}=\pi$, the analytical expression of $\Delta{\zeta}^{S_{x}}$ is provided as
\begin{equation}
	\Delta{\zeta}_{\Phi=\pi}^{S_{x}}=\frac{\sqrt{2}\lbrack{3-2\cos(\frac{\theta_{1}+\theta_{2}}{2})+\cos(\theta_{1}+\theta_{2})}\rbrack}{\sqrt{3}\lbrack{9\eta^{'''}-2\delta^{'''}-18\lambda^{'''}-36\gamma^{'''}+26\varpi^{'''}+14\varepsilon^{'''}+\upsilon^{'''}+29}\rbrack^{\frac{1}{2}}},
	\label{35}
\end{equation}
where the coefficients are 
\begin{align}
	&\eta^{'''}=\cos(2\theta_{1})+\cos(2\theta_{2}),\hspace{1cm}
	\delta^{'''}=\cos\bigl(\frac{5\theta_{1}+\theta_{2}}{2}\bigl)+\cos\bigl(\frac{\theta_{1}+5\theta_{2}}{2}\bigl)+2\cos\bigl(\frac{3}{2}(\theta_{1}+\theta_{2})\bigl),\hspace{1cm}\varepsilon^{'''}=\cos(\theta_{1}+\theta_{2}),\notag\\&
	\lambda^{'''}=\cos\bigl(\frac{1}{2}(3\theta_{1}-\theta_{2}))+\cos(\frac{1}{2}(\theta_{1}-3\theta_{2})\bigl),\hspace{0.5cm}\gamma^{'''}=\cos\bigl(\frac{1}{2}(\theta_{1}+\theta_{2})\bigl),\hspace{0.5cm}\varpi^{'''}=\cos(\theta_{1}-\theta_{2}),\hspace{1cm}\upsilon^{'''}=\cos(2(\theta_{1}+\theta_{2})).\label{36}
\end{align}
\section{Quantum Fisher information for the $s$-spin coherent states}\label{AppC}
For the $s$-spin coherent states of the form \eqref{44}, one verifies that the quantum Fisher information \eqref{6} can be represented as
\begin{align}
	F(\rho_{\zeta},S_{z})&=4\left\lbrace {\cal N}_{c}^{2}\sum_{m=-s}^{s}\frac{2s!m^{2}}{(s+m)!(s-m)!}\Biggr[{\frac{|z_{1}|^{2(s+m)}}{(1+|z_{1}|^{2})^{2s}}+\frac{|z_{2}|^{2(s+m)}}{(1+|z_{2}|^{2})^{2s}}+\frac{\bar{z_{1}}^{s+m}{z_{2}^{s+m}}+\bar{z_{2}}^{s+m}{z_{1}}^{s+m}}{(1+|z_{1}|^{2})^{s}(1+|z_{2}|^{2})^{s}}}\Biggr]\right.\notag\\&\left. -\Biggr|{\cal N}_{c}^{2}\sum_{m=-s}^{s}\frac{2s!m}{(s+m)!(s-m)!}\Biggr[{\frac{|z_{1}|^{2(s+m)}}{(1+|z_{1}|^{2})^{2s}}+\frac{|z_{2}|^{2(s+m)}}{(1+|z_{2}|^{2})^{2s}}+\frac{\bar{z_{1}}^{s+m}{z_{2}}^{s+m}+\bar{z_{2}}^{s+m}{z_{1}}^{s+m}}{(1+|z_{1}|^{2})^{s}(1+|z_{2}|^{2})^{s}}}\Biggr]\Biggr|^{2}\right\rbrace,\label{48}
\end{align}
\begin{align}
	F(&\rho_{\zeta},S_{x})=4\left\lbrace\frac{{\cal N}_{c}^{2}}{4}\left[ 2s(2s-1)\left( \frac{\bar{z_{1}}^{2}+{z_{1}}^{2}}{(1+|z_{1}|^{2})}+\frac{\bar{z_{2}}^{2}+{z_{2}}^{2}}{(1+|z_{2}|^{2})}\right) +\frac{2s(2s-1)}{(1+|z_{1}|^{2})^{s}(1+|z_{2}|^{2})^{s}}\times \right.\right.  \notag\\&\left.\left.\left(\frac{(\bar{z_{1}}^{2}+{z_{2}}^{2})(1+\Bar{z_{1}}{z_{2}})^{2s}}{(1+\Bar{z_{1}}{z_{2}})^{2}}+\frac{(\bar{z_{2}}^{2}+{z_{1}}^{2})(1+\Bar{z_{2}}{z_{1}})^{2s}}{(1+\Bar{z_{2}}{z_{1}})^{2}}\right)+\frac{8s^{2}\bar{z_{1}}z_{1}+2s(\bar{z_{1}}z_{1})^{2}+2s}{(1+\bar{z_{1}}z_{1})^{2}}+\frac{8s^{2}\bar{z_{2}}z_{2}+2s(\bar{z_{2}}z_{2})^{2}+2s}{(1+\bar{z_{2}}z_{2})^{2}}\right.\right. \notag\\&\left.\left. +\frac{1}{(1+|z_{1}|^{2})^{s}(1+|z_{2}|^{2})^{s}}\left({\frac{(1+\Bar{z_{1}}{z_{2}})^{2s}}{(1+\Bar{z_{1}}{z_{2}})^{2}}\Bigl(8s^{2}\Bar{z_{1}}{z_{2}}+2s(\Bar{z_{1}}{z_{2}})^{2}+2s\Bigl)}+
	{\frac{(1+\Bar{z_{2}}{z_{1}})^{2s}}{(1+\Bar{z_{2}}{z_{1}})^{2}}\Bigl(8s^{2}\Bar{z_{2}}{z_{1}}+2s(\Bar{z_{2}}{z_{1}})^{2}+2s\Bigl)}\right)\right] \right.\notag\\&\left.-\Bigg|\frac{{\cal N}_{c}^{2}}{2}\Biggr[2s\left({\frac{\bar{z_{1}}+{z_{1}}}{(1+|z_{1}|^{2})}}+{\frac{\bar{z_{2}}+{z_{2}}}{(1+|z_{2}|^{2})}}\right)+
	\frac{2s}{(1+|z_{1}|^{2})^{s}(1+|z_{2}|^{2})^{s}}
	\left( \frac{(\bar{z_{1}}+{z_{2}})(1+\bar{z_{1}}{z_{2}})^{2s}}{1+\bar{z_{1}}{z_{2}}}+ \frac{(\bar{z_{2}}+{z_{1}})(1+\bar{z_{2}}{z_{1}})^{2s}}{1+\bar{z_{2}}{z_{1}}}\right)\Biggr]\Bigg|^{2}\right\rbrace,
\end{align}
and
\begin{multline}
	F(\rho_{\zeta},S_{y})=4\Biggl\{\frac{-{\cal N}_{c}^{2}}{4}\Biggr[2s(2s-1)~\Biggl(\frac{\bar{z_{1}}^{2}+{z_{1}}^{2}}{(1+|z_{1}|^{2})}+\frac{\bar{z_{2}}^{2}+{z_{2}}^{2}}{(1+|z_{2}|^{2})}\Biggl)+\frac{2s(2s-1)}{(1+|z_{1}|^{2})^{s}(1+|z_{2}|^{2})^{s}}\times\\
	\Biggl(\frac{(\bar{z_{1}}^{2}+{z_{2}}^{2})(1+\Bar{z_{1}}{z_{2}})^{2s}}{(1+\Bar{z_{1}}{z_{2}})^{2}}+\frac{(\bar{z_{2}}^{2}+{z_{1}}^{2})(1+\Bar{z_{2}}{z_{1}})^{2s}}{(1+\Bar{z_{2}}{z_{1}})^{2}}
	\Biggl)-\frac{8s^{2}\bar{z_{1}}z_{1}+2s(\bar{z_{1}}z_{1})^{2}+2s}{(1+\bar{z_{1}}z_{1})^{2}}-
	\frac{8s^{2}\bar{z_{2}}z_{2}+2s(\bar{z_{2}}z_{2})^{2}+2s}{(1+\bar{z_{2}}z_{2})^{2}}\\-\frac{1}{(1+|z_{1}|^{2})^{s}(1+|z_{2}|^{2})^{s}}\Biggl({\frac{(1+\Bar{z_{1}}{z_{2}})^{2s}}{(1+\Bar{z_{1}}{z_{2}})^{2}}\Bigl(8s^{2}\Bar{z_{1}}{z_{2}}+2s(\Bar{z_{1}}{z_{2}})^{2}+2s\Bigl)}+
	{\frac{(1+\Bar{z_{2}}{z_{1}})^{2s}}{(1+\Bar{z_{2}}{z_{1}})^{2}}\Bigl(8s^{2}\Bar{z_{2}}{z_{1}}+2s(\Bar{z_{2}}{z_{1}})^{2}+2s\Bigl)}\Biggl)\Biggr]\\-\Bigg|\frac{{\cal N}_{c}^{2}}{2i}\Biggr[2s\Biggl({\frac{\bar{z_{1}}-{z_{1}}}{(1+|z_{1}|^{2})}}+{\frac{\bar{z_{2}}-{z_{2}}}{(1+|z_{2}|^{2})}}\Biggl)+
	\frac{2s}{(1+|z_{1}|^{2})^{s}(1+|z_{2}|^{2})^{s}}
	\Biggl(\frac{(\bar{z_{1}}-{z_{2}})(1+\bar{z_{1}}{z_{2}})^{2s}}{1+\bar{z_{1}}{z_{2}}}+ \frac{(\bar{z_{2}}-{z_{1}})(1+\bar{z_{2}}{z_{1}})^{2s}}{1+\bar{z_{2}}{z_{1}}}\Biggl)\Biggr]\Bigg|^{2}\Biggl\},
\end{multline}
for $H=S_{z}$, $H=S_{x}$ and $H=S_{y}$, respectively.\\

{\bf Contributions:}\par
All authors contributed equally to the paper.

\end{document}